\documentclass[lettersize,journal]{IEEEtran}
\usepackage{array}
\usepackage[caption=false,font=normalsize,labelfont=sf,textfont=sf]{subfig}
\usepackage{stfloats}
\usepackage{verbatim}

\usepackage{cite}
\usepackage{bm}
\usepackage{amsmath,amssymb,amsfonts, amsthm}
\usepackage{algorithm}
\usepackage{algorithmic}
\usepackage{graphicx}
\usepackage{textcomp}
\usepackage{xcolor}
\usepackage{gensymb}
\usepackage{comment}
\usepackage{soul}
\usepackage{color, xcolor}
\usepackage{amsthm}
\usepackage{bm}
\usepackage{multicol}
\usepackage{enumitem}
\usepackage{booktabs}
\usepackage{array}
\usepackage{textcomp}
\allowdisplaybreaks[4]

\newtheorem{definition}{Definition}
\newtheorem{lemma}{Lemma}
\newtheorem{theorem}{Theorem}
\newtheorem{remark}{Remark}
\newcolumntype{C}[1]{>{\centering\arraybackslash}p{#1}}
\newcolumntype{M}[1]{>{\centering\arraybackslash}m{#1}}

\def\BibTeX{{\rm B\kern-.05em{\sc i\kern-.025em b}\kern-.08em
		T\kern-.1667em\lower.7ex\hbox{E}\kern-.125emX}}
\usepackage{balance}

\begin{document}
	\title{Robust Multi-Stream Massive MIMO Satellite Systems Based on Statistical CSI}
	
	\author{Hangsong Yan, \textit{Member}, \textit{IEEE}, Alexei Ashikhmin, \textit{Fellow}, \textit{IEEE}, Hong Yang, \textit{Senior Member}, \textit{IEEE}, Bin Song, \textit{Senior Member}, \textit{IEEE}, and Shu Sun, \textit{Senior Member}, \textit{IEEE}
		\thanks{
			Hangsong Yan and Bin Song are with the Hangzhou Institute of Technology, Xidian University, China (e-mails: yanhangsong@xidian.edu.cn, bsong@mail.xidian.edu.cn). 
			
			Alexei Ashikhmin is with the Department of Mathematics and Algorithms Research, Nokia Bell Labs, USA (e-mail: alexei.ashikhmin@nokia-bell-labs.com). Hong Yang (retired) was with the Department of Mathematics and Algorithms Research, Nokia Bell Labs, USA, and is now retired (e-mail: hyang.bell.labs@gmail.com).
			
			Shu Sun is with the School of Information Science and Electronic Engineering, Shanghai Jiao Tong University, Shanghai, China (e-mail: shusun@sjtu.edu.cn).
			
			
			Part of this work was accepted by IEEE International Conference on Communications Workshops, Montreal, Canada, June 2025 \cite{Yan2025Robust}.
	}}
	
	
	\maketitle
	
	\begin{abstract}
		This paper investigates multi-stream downlink precoding for massive multiple-input multiple-output low-Earth-orbit satellite (SAT) communication systems. We adopt a delay and Doppler precompensation approach to achieve coherent transmission. Under this setting, we formulate a signal transmission model that incorporates the near-independent properties of inter-SAT interference and compensation errors. We then demonstrate that moving beyond single-stream transmission requires both multi-SAT cooperation and multi-antenna user terminals. Based on this configuration and the established signal transmission model, we derive the first- and second-order statistical channel characteristics and utilize them to design locally optimal precoding algorithms for both total power constraint (TPC) and per-antenna power constraint (PAPC) conditions, which rely only on statistical channel state information (sCSI). In particular, the designed PAPC algorithm achieves linear complexity with respect to the number of antennas on the cooperative SATs. To reduce the computational complexity of the locally optimal precoder under TPC, we propose a low-complexity and robust precoding scheme targeting both the minimum mean squared error and sum-rate maximization objectives. Using majorization theory, we also provide a rigorous theoretical analysis of the optimal precoding structure under TPC. Moreover, the Lanczos algorithm is adopted to further reduce the complexity of the proposed robust designs. Simulation results show that when each SAT is equipped with a sufficiently large number of antennas, the proposed sCSI-based designs achieve performance comparable to that of instantaneous CSI-based designs.
	\end{abstract}
	

	\section{Introduction}
	\IEEEPARstart{U}{biquitous} seamless global coverage is regarded as a key vision in the 6G era \cite{ChengXiang2023Road}. Moreover, the demand for higher throughput in mobile communication systems has consistently driven the evolution of networks from the first generation toward 5G and beyond. Satellite (SAT) communication possesses the distinctive ability to cover wide geographical areas especially in areas underserved by terrestrial mobile networks\cite{38.811_3GPP, Jamshed2025Tutorial}. In recent years, broadband low-Earth-orbit (LEO) SAT communication systems have attracted increased interest from both academia \cite{Qu2017LEO, Di2019Ultra} and industry \cite{38.811_3GPP, DELPORTILLO2019Comparison}. This is due to their inherent advantages over traditional geostationary-Earth-orbit SAT systems, including shorter round-trip delays, reduced path loss, and lower launch costs. 
	Furthermore, LEO systems are better suited for direct SAT-to-ground communication, particularly for applications requiring high-speed data transmission. 
	
	However, LEO SAT communication continues to face multiple challenges. The long distance between LEO SATs and ground user terminals (UTs) causes higher path loss compared to terrestrial communication. Furthermore, the high-speed motion of LEO SATs not only causes large Doppler shifts but also leads to severe channel aging, making it difficult to acquire accurate instantaneous channel state information (iCSI) at the SATs. To tackle these challenges, massive multiple-input multiple-output (mMIMO) has been identified as a promising enabler\cite{Marzetta2010NonCo, Larsson2014mMIMO}. Its ``favorable propagation'' and ``channel hardening'' properties\cite{Marzetta2016Fundamentals} not only provide high beamforming gain to combat high path loss but also make precoding based on statistical CSI (sCSI) possible.
	
	Motivated by the aforementioned advantages of mMIMO technology, the authors of \cite{You2020mMIMO, Ke-Xin2022Downlink, Xiang2024Massive} conducted a series of studies applying mMIMO to LEO SAT communication systems. 
	Based on a detailed analysis of SAT communication channels, they concluded that when a ground UT is served by a single SAT, single-stream data transmission is the optimal method for that UT, regardless of whether it is equipped with one or multiple antennas. Building on this analysis, they further proposed a robust sCSI-based precoding scheme and a strategy for SAT-UT association \cite{You2020mMIMO}. Additionally, they investigated single-stream transmission for a single UT under multi-SAT cooperation. For this scenario, they proposed a joint precoding and receiver design based on the weighted minimum mean squared error (WMMSE) principle that can utilize both iCSI and sCSI \cite{Xiang2024Massive}. Furthermore, \cite{Ziyu2026Decoupled} investigated the decoupled design of precoders and receivers for both downlink (DL) and uplink (UL) transmissions in LEO multi-SAT communication systems. By establishing a beam-based SAT channel model, they demonstrated that the design of the space-domain DL precoder and UL receiver can be converted into that of lower-dimensional beam-domain vectors with local sCSI. Exploiting the properties of the beam matrix, the authors proposed low-complexity decoupled designs for the DL precoder and UL receiver.
	
	Regarding the foundational architecture of LEO SAT mMIMO communication, the authors of \cite{Zhao2023Robust} proposed a robust DL transmission design assisted by intelligent reflecting surfaces in cognitive SAT and terrestrial networks. By employing angle discretization and successive convex approximation methods, they solved the total transmission power minimization problem in this scenario, thereby achieving optimal energy efficiency. To reduce computational complexity, they also proposed a low-complexity algorithm combining generalized zero-forcing beamforming and alternating optimization. A joint hybrid beamforming and user scheduling design for multi-SAT cooperative networks was proposed in \cite{Zhang2024Multi}. By exploring the intrinsic connection between beamforming and user scheduling, this design significantly improved network performance. Multi-SAT cooperation was also considered in \cite{Guidotti2024Federated}, where it was referred to as a federated cooperative cell-free MIMO system. This research provided a detailed discussion of the architecture and performance of this system, proposing a novel location-based cell-free MIMO algorithm and associated power distribution algorithms.
	
	We note that most recent mMIMO SAT communication studies, including those previously mentioned, have focused on scenarios with either a single antenna at the UT or single-stream data transmission per UT. Such scenarios are typically well-suited for applications such as the Internet of Things (IoT). The study in \cite{Abdelsadek2023Broadband} investigated multi-SAT cooperation, allowing a single UT to receive multiple data streams. This work adopted a conjugate beamforming (CB) precoding scheme. By combining this scheme with the characteristics of SAT communication channels, the authors derived a closed-form expression for spectral efficiency (SE) that depends only on large-scale channel coefficients. They also conducted a systematic performance comparison, highlighting the advantages of multi-SAT cooperation over single-SAT services. However, the CB scheme evaluated in \cite{Abdelsadek2023Broadband} relies on iCSI, and CB itself has limited ability to suppress inter-stream and inter-UT interference; consequently, the precoding scheme requires further optimization. Furthermore, the effect of asynchronous reception caused by large propagation delay differences in multi-SAT systems \cite{Yan2019Asyn} was not accounted for. In \cite{Yafei2026Statistical}, the authors introduced delay and Doppler precompensation to ensure coherent transmission in cooperative SAT systems. Under this setting, inter-SAT interference exhibits a near-independent property. Accounting for this property and the compensation errors, the authors derived a covariance decomposition-based WMMSE formulation for precoding design that depends on sCSI. They also proposed a model-driven, scalable deep learning approach to reduce computational complexity. However, \cite{Yafei2026Statistical} focuses on single-stream transmission for each UT and does not address the more practical per-antenna power constraint (PAPC) requirements.
	
	In this work, we focus on the DL precoding design for an S-band\footnote{This band is for ground handheld UTs in LEO SAT communications\cite{38.811_3GPP}.} LEO multi-SAT communication system incorporating delay and Doppler precompensation.
	Each SAT is equipped with a large number of antennas and each UT has multiple antennas, enabling multi-stream transmission. Multiple UTs are served by more than one SAT using the same time and frequency resources. To the best of our knowledge, such a scenario utilizing sCSI-based precoding has not been treated before. We stress that this multi-antenna UT configuration is prevalent, since a large portion of handheld devices (e.g., smartphones) are now equipped with multiple antennas. However, the considered scenario faces the following challenges: a) The long propagation delay of SAT communication leads to rapid aging of iCSI, requiring a robust transmission framework based on sCSI, which is required to be adaptable to long-term channel characteristics. b) In addition to enhancing the power of desired signals, the precoder design needs to suppress interference between UTs and among multiple data streams within each UT, leading to high algorithmic complexity. 
	c) The PAPC is more practical but imposes more stringent limitations compared to the TPC, making it difficult to obtain a solution that balances performance and algorithmic complexity. 
	To tackle these challenges, the main contributions of this work are summarized as follows:
	
	1) We adopt a delay and Doppler precompensation approach to ensure coherent transmission and formulate a multi-stream transmission model that incorporates the near-independent properties of inter-SAT interference and compensation errors.
	
	2) We perform a detailed analysis of SAT communication channel characteristics, demonstrating that multi-stream transmission to a single UT requires both multi-SAT cooperation and multi-antenna UTs. By exploiting the Rician-distributed channel characteristics, we derive the first- and second-order channel statistics.  
	
	3) Based on the WMMSE framework, we provide locally optimal iCSI-based precoding algorithms for both TPC and PAPC conditions. In particular, by sequentially updating the precoding vector of each antenna, the designed PAPC algorithm achieves linear computational complexity with respect to the number of antennas on the cooperative SATs.
	
	4) By utilizing the obtained first- and second-order channel statistics, we derive statistical expressions for the achievable rate, mean squared error (MSE) matrix, and MMSE matrix. Subsequently, we propose locally optimal sCSI-based precoding designs under the WMMSE framework for both TPC and PAPC.
	
	5) To tackle the high computational complexity (i.e., cubic with respect to the number of antennas on the cooperative SATs) in the TPC design, we propose a robust and low-complexity precoding scheme leveraging sCSI that targets both the MMSE and sum-rate maximization criteria. Using majorization theory, we provide a rigorous theoretical analysis of the optimal precoding structure under TPC in Theorem \ref{theorem:eigenvalues}. To the best of our knowledge, such a rigorous treatment based on majorization theory has not been presented before.
	
	6) We employ the Lanczos algorithm to further reduce the computational complexity of the eigenvalue decomposition (EVD) in the proposed low-complexity precoding scheme, with only a slight performance degradation. Comprehensive simulations are conducted, and the results show that the performance of the proposed sCSI-based precoding schemes is comparable to that of the locally optimal iCSI-based precoding scheme when each SAT is equipped with a sufficiently large number of antennas. 
	
	The remainder of this paper is organized as follows. Section \ref{sec:System_Model} presents the multi-stream system model, including the channel model, the delay and Doppler precompensation procedure, and the statistical SAT channel properties. Section \ref{sec:optimal_design} details the locally optimal iCSI-based precoder designs under the WMMSE framework. Section \ref{sec:Optimal precoder design based on sCSI} develops the corresponding locally optimal precoder designs based on sCSI.
	Section \ref{sec:Robust Precoding Design with TPC} proposes robust and low-complexity precoding algorithms under TPC. Furthermore, this section provides a rigorous theoretical analysis of the optimal precoding matrix structure and details the application of the Lanczos algorithm for reducing EVD complexity.
	Finally, Section \ref{sec:simulations} evaluates the proposed schemes through extensive simulations, and Section \ref{sec:conclusion} concludes the paper.
	
	Notations: Lowercase letters denote scalars, while boldface lowercase (uppercase) letters denote vectors (matrices), respectively. $\text{tr}(\cdot)$, $\text{rank}(\cdot)$, $(\cdot)^{*}$, $(\cdot)^T$, and $(\cdot)^H$ denote the trace, rank, conjugate, transpose, and conjugate transpose of a matrix, respectively. The symbol $\otimes$ denotes the Kronecker product. 
	$[\mathbf{x}]_i$ denotes the $i$-th element of vector $\mathbf{x}$, $[\mathbf{X}]_{i}$ denotes the $i$-th column of matrix $\mathbf{X}$, and $[\mathbf{X}]_{i,j}$ denotes the element of matrix $\mathbf{X}$ in the $i$-th row and $j$-th column. $\text{diagblk}\{[\mathbf{X}_1, \dots,\mathbf{X}_N]\}$ denotes a block diagonal matrix with diagonal blocks $[\mathbf{X}_1, \dots, \mathbf{X}_N]$. $\text{diag}(\mathbf{X})$ denotes a vector consisting of the diagonal elements of $\mathbf{X}$, whereas $\text{diag}(\mathbf{x})$ denotes a diagonal matrix whose diagonal elements are formed by the entries of vector $\mathbf{x}$.
	
	\section{System Model}
	\label{sec:System_Model}
	We consider DL transmission in an LEO SAT mMIMO communication system, where SATs are grouped into clusters. In each cluster, the SATs are connected via inter-SAT links (ISLs) to a central processing unit deployed on a super-SAT \cite{Abdelsadek2023Broadband}. This super-SAT controls the cooperative transmission from these SATs to the UTs. Specifically, we assume that $S$ SATs within each cluster cooperate to serve $K$ ground UTs using the same time and frequency resource blocks. 
	Each SAT is equipped with an $M$-antenna uniform planar array (UPA) with $M_x$ antennas along the $x$-axis and $M_y$ antennas along the $y$-axis. Each UT is equipped with an $N$-antenna UPA with $N_{x'}$ antennas along the $x'$-axis and $N_{y'}$ antennas along the $y'$-axis. For each SAT, the $z$-axis of the UPA points toward the Earth's center, while the $x$- and $y$-axes are oriented toward the geographic South Pole and the west, respectively. For each UT, we assume that its UPA lies on the horizontal $x'-y'$ plane and that the $z'$-axis is perpendicular to the ground.
	We also note that free-space optical technologies \cite{Chaudhry2021Free} can support high-speed, reliable, and low-latency ISLs, helping mitigate fronthaul signaling overhead issues.
	
	\subsection{Channel Model}
	\label{sec:Signal and Channel Model}
	We model the time-frequency baseband channel from the $s$-th SAT to the $k$-th UT for the signal radiated at time $t$ as:
		\begin{equation} \label{eq:H_k_s_t_f}
			\tilde{\mathbf{H}}_{s,k}(t, f) = \sum_{l=1}^{L_{s,k}}a_{s,k,l}e^{j2\pi [v_{s,k,l}t - f\tau_{s,k,l}]}\mathbf{d}_{s,k,l}\mathbf{g}_{s,k,l}^{H}, 
	\end{equation}
	where $j \triangleq \sqrt{-1}$, and $L_{s,k}$ is the number of channel paths from SAT $s$ to UT $k$. For the $l$-th path, $a_{s,k,l}$, $v_{s,k,l}$, and $\tau_{s,k,l}$ are the complex channel gain, Doppler shift, and propagation delay, respectively, while $\mathbf{d}_{s,k,l}\in \mathbb{C}^{N\times 1}$ and $\mathbf{g}_{s,k,l}\in\mathbb{C}^{M \times 1}$ are the array response vectors at the UT and SAT sides, respectively. In SAT communications, scatterers near the SAT are typically absent, and the distance between the SAT and the UT is very large. Thus, it is reasonable to assume that a single transmission path exists at the SAT side \cite{38.811_3GPP}. Under this assumption, the SAT-side array response vector is the same for all propagation paths at the UT side (i.e., $\mathbf{g}_{s,k,l} = \mathbf{g}_{s,k}, \forall l$). 
	$\mathbf{g}_{s,k}$ and $\mathbf{d}_{s,k,l}$ can be expressed as follows:
		\begin{equation}
			\mathbf{g}_{s,k} = \mathbf{v}_{x}(\cos\theta_{s,k}^x\sin\theta_{s,k}^y) \otimes \mathbf{v}_{y}(\cos\theta_{s,k}^y),
		\end{equation}
		\begin{equation}
			\mathbf{d}_{s,k,l} = \mathbf{v}_{x'}(\cos\theta_{s,k,l}^{x'}\sin\theta_{s,k,l}^{y'}) \otimes \mathbf{v}_{y'}(\cos\theta_{s,k,l}^{y'}),
		\end{equation}
		where $\theta^x$, $\theta^y$, $\theta^{x'}$, and $\theta^{y'}$ denote the angles with respect to the $x$, $y$, $x'$, and $y'$ axes, respectively. Letting $\upsilon \in \{x, y, x', y'\}$, $\mathbf{v}_\upsilon(x)$ is defined by:
		\begin{equation}
			\mathbf{v}_\upsilon(x) = \frac{1}{\sqrt{M_\upsilon}}[1, e^{-j2\pi\frac{d_{\upsilon}}{\lambda}x},\dots, e^{-j2\pi(M_\upsilon - 1)\frac{d_\upsilon}{\lambda}x}]^T,
		\end{equation} 
		where $M_{\upsilon}$ is the number of antennas along the $\upsilon$-axis, $d_\upsilon$ is the antenna spacing along the $\upsilon$-axis, and $\lambda$ is the operating wavelength.
		The Doppler shift $v_{s,k,l}$ also consists of contributions from both the SAT and the UT sides. The Doppler shift caused by the SAT motion is assumed to be the same for all multipath components. Specifically, we have $v_{s,k,l} = v_{s,k}^{\text{sat}} + v_{s,k,l}^{\text{UT}}$. Here, we assume that $v_{s,k,l}^{\text{UT}}$ has been properly compensated for and has a negligible effect on system performance. We also model the propagation delay as $\tau_{s,k,l} = \tau_{s,k}^{\text{min}} + \tau_{s,k,l}^{\text{UT}}$, where $\tau_{s,k}^{\text{min}}$ is the propagation delay associated with the line-of-sight (LoS) path between the $s$-th SAT and the $k$-th UT. Based on the above description, $\tilde{\mathbf{H}}_{s,k}(t, f)$ can be rewritten as:
		\begin{equation}
			\tilde{\mathbf{H}}_{s,k}(t, f) = e^{j2\pi[v_{s,k}^{\text{sat}}t - f\tau_{s,k}^{\text{min}}]}\mathbf{d}_{s,k}(f)\mathbf{g}_{s,k}^H,
		\end{equation}
		where	
		\begin{equation}
			\mathbf{d}_{s,k}(f) = \sum_{l=1}^{L_{s,k}}a_{s,k,l}e^{-j2\pi f\tau_{s,k,l}^{\text{UT}}}\mathbf{d}_{s,k,l}.
	\end{equation}
	
	\subsection{Transmitter Delay and Doppler Precompensation}
		Let $\mathbf{x}_{s,k,r}^{(q)}$ denote the signal transmitted from the $s$-th SAT to the $k$-th UT on the $r$-th subcarrier of the $q$-th OFDM symbol.
		By applying the inverse Fourier transform to $\mathbf{x}_{s,k,r}^{(q)}$, the time-domain baseband signal is given by:
		\begin{equation}
			\mathbf{x}_{s,k}(t) = \sum_{r=0}^{N_r-1}\sum_{q=0}^{N_q-1}\mathbf{x}_{s,k,r}^{(q)}u(t-qT_{\text{sym}})e^{j2\pi r \Delta f(t - qT_{\text{sym}})}.
		\end{equation}
		Here, $T_{\text{sym}} = T + T_{\text{CP}}$, where $T_{\text{sym}}$, $T$, and $T_{\text{CP}}$ denote the OFDM symbol duration, data duration, and cyclic prefix (CP) duration, respectively. The function $u(t) = 1$ if $t \in [0, T]$ and $u(t) = 0$ otherwise. Additionally, $\Delta f = \frac{1}{T}$ is the subcarrier spacing, while $N_r$ and $N_q$ denote the number of subcarriers and consecutive OFDM symbols, respectively.
		After appending the CP and applying frequency upconversion, the transmitted signal can be written as:
		\begin{equation}
			\hat{\mathbf{x}}_{s,k}(t) = \left\{ \begin{array}{rcl}
				e^{j2\pi f_0 t}\mathbf{x}_{s,k}(t+T), &  t \in [qT_{\text{sym}} - T_{\text{CP}}, qT_{\text{sym}}], \\ 
				e^{j2\pi f_0 t}\mathbf{x}_{s,k}(t), & t \in [qT_{\text{sym}}, qT_{\text{sym}} + T],
			\end{array}\right.
		\end{equation}
		where $f_0$ denotes the carrier frequency. To achieve coherent transmission, we adopt the compensation approach utilized in \cite{Yafei2026Statistical}. By applying the delay and Doppler precompensation offsets, denoted by $\tau_{s,k}^{\text{cps}}$ and $v_{s,k}^{\text{cps}}$, respectively, to $\hat{\mathbf{x}}_{s,k}(t)$, we obtain:
		\begin{equation}
			\hat{\mathbf{x}}_{s,k}^{\text{cps}}(t) = \hat{\mathbf{x}}_{s,k}(t + \tau_{s,k}^{\text{cps}})e^{-j2\pi(t + \tau_{s,k}^{\text{cps}})v_{s,k}^{\text{cps}}}.
	\end{equation}
	
	At the UT, the compensated time-domain signal is convolved with the channel impulse response, followed by downconversion and a discrete Fourier transform. The received signal at the $k$-th UT on the $r$-th subcarrier during the $q$-th OFDM symbol duration can be expressed as\footnote{Interested readers are referred to \cite{Yafei2026Statistical} for the detailed mathematical derivation of this procedure.}:
	\begin{align}
		\mathbf{y}_{k, r}^{(q)} = \sum_{s=1}^{S}\mathbf{H}_{s,k,r}\Big(\mathbf{x}_{s,k,r}^{(q)}\varphi_{s,k,r}^{(q)} + \sum_{k'\neq k}^{K}\breve{\mathbf{x}}_{s,k',r}^{(q'_{s,k',k})}\varphi_{s,k',k,r}^{(q)}\Big),
		\label{eq:y_k_r_m}
	\end{align}
	where $\mathbf{H}_{s,k,r}$ is the channel frequency response at the $r$-th subcarrier, given by:
	\begin{equation}
		\mathbf{H}_{s,k,r} = \sum_{l=1}^{L_{s,k}} a_{s,k,l}e^{-j2\pi \frac{r}{T}\tau_{s,k,l}^{\text{UT}}}\mathbf{d}_{s,k,l}\mathbf{g}_{s,k}^{H}.
		\label{eq:H_s_k_r}
	\end{equation}
	The term $\varphi_{s,k,r}^{(q)}$ is the residual phase error due to precompensation inaccuracies and is defined as:
	\begin{equation}
		\varphi_{s,k,r}^{(q)} = e^{j2\pi[(f_0 + r \Delta f - v_{s,k}^{\text{cps}})\bar{\tau}_{s,k} + qT_{\text{sym}}\bar{v}_{s,k}]},
		\label{eq:phase_error}
	\end{equation}
	where $\bar{\tau}_{s,k}$ and $\bar{v}_{s,k}$ denote the delay and Doppler compensation errors, respectively, which are defined as:
	\begin{equation}
		\bar{\tau}_{s,k} = \tau_{s,k}^{\text{cps}} - \tau_{s,k}^{\text{min}},\quad \bar{v}_{s,k} = v_{s,k}^{\text{sat}} - v_{s,k}^{\text{cps}}.
	\end{equation}
	Additionally, $\varphi_{s,k',k,r}^{(q)}$ denotes the phase difference between the compensation for the $k'$-th UT and the delay and Doppler shifts associated with the $k$-th UT. Its expression is given by:
	\begin{equation}
		\varphi_{s,k',k,r}^{(q)} = e^{j2\pi[(f_0 + r\Delta f - v_{s,k'}^{\text{cps}})\tilde{\tau}_{s,k',k} + qT_{\text{sym}}\tilde{v}_{s,k',k}]},
	\end{equation}
	where
	\begin{equation}
		\tilde{\tau}_{s,k',k} = \tau_{s,k'}^{\text{cps}} - \tau_{s,k}^{\text{min}}, \quad \tilde{v}_{s,k',k} = v_{s,k}^{\text{sat}} - v_{s,k'}^{\text{cps}}.
	\end{equation}
	The term $\breve{\mathbf{x}}_{s,k',r}^{(q'_{s,k',k})}$ denotes the signal transmitted from the $s$-th SAT to the $k'$-th UT, but demodulated at the $k$-th UT. Due to the potentially large values of $|\tilde{\tau}_{s,k',k}|$ and $|\tilde{v}_{s,k',k}|$,  $\breve{\mathbf{x}}_{s,k',r}^{(q'_{s,k',k})}$ typically consists of components from multiple subcarriers, namely $\mathbf{x}_{s,k',1}^{(q'_{s,k',k})},\dots ,\mathbf{x}_{s,k',N_r}^{(q'_{s,k',k})}$. Furthermore, it may contain information from more than one OFDM symbol; thus, we use $q'_{s,k',k}$ to denote an index spanning more than one symbol.
	\begin{remark}
		A significant difference between cooperative non-terrestrial communications and terrestrial communications is that the delay difference between interference signals from different SATs is larger than the OFDM symbol duration. Specifically, we have $\Delta\tau_{k',k}^{s_1, s_2} = |(\tau_{s_1, k'}^{\text{min}} - \tau_{s_1, k}^{\text{min}}) - (\tau_{s_2, k'}^{\text{min}} - \tau_{s_2, k}^{\text{min}})| \approx |\tilde{\tau}_{s_1, k',k} - \tilde{\tau}_{s_2, k',k}| > T_{\text{sym}}$. Therefore, we can assume that $\breve{\mathbf{x}}_{s_1, k',r}^{(q'_{s_1, k',k})}$ and $\breve{\mathbf{x}}_{s_2, k',r}^{(q'_{s_2, k',k})}$ contain different symbol information and can be regarded as statistically independent \cite{Yafei2026Statistical}.
		\label{remark:interference_independent}
	\end{remark}
	Since the same signal processing approach is applied across different time-frequency resource blocks, we omit the subcarrier index $r$ and the OFDM symbol index $q$ hereafter for brevity.
	Consequently, the received signal at the $k$-th UT can be expressed as:
	\begin{align}
		\mathbf{y}_{k} = \sum_{s=1}^{S}\mathbf{H}_{s,k}\Big(\mathbf{W}_{s,k}\mathbf{s}_{k}\varphi_{s,k} + \sum_{k'\neq k}^{K}\mathbf{W}_{s,k'}\breve{\mathbf{s}}_{s,k'}\varphi_{s,k',k}\Big) + \mathbf{n}_{k},
		\label{eq:y_k}
	\end{align}
	where $\mathbf{s}_k \in \mathbb{C}^{d_k \times 1}$ denotes the transmitted multi-stream signal for the $k$-th UT, with $d_k$ representing the number of signal streams\footnote{In FDD mode, $d_k$ is fed back by the UTs.}. Additionally, $\breve{\mathbf{s}}_{s,k'}$ is the signal information contained in $\breve{\mathbf{x}}_{s,k'}$, $\mathbf{W}_{s,k} \in \mathbb{C}^{M \times d_k}$ is the precoding matrix at the $s$-th SAT for the $k$-th UT, and $\mathbf{n}_k \sim \mathcal{CN}(\mathbf{0}, \sigma_{k}^2 \mathbf{I}_N)$ is the noise vector at the $k$-th UT. 
	
	\subsection{Statistical SAT Channel Properties}
	\label{sec:statistical_channel_model}
	Let $\hat{\mathbf{H}}_{s,k} = \mathbf{H}_{s,k}\varphi_{s,k}$ and $\hat{\mathbf{H}}_{k} = [\hat{\mathbf{H}}_{1,k}\, \hat{\mathbf{H}}_{2,k}\, \cdots\, \hat{\mathbf{H}}_{S,k}]\in \mathbb{C}^{N\times SM}$ represent the compensated channel from the $S$ SATs to the $k$-th UT. 
	Based on (\ref{eq:H_s_k_r}), $\hat{\mathbf{H}}_k$ can be rewritten as:
	\begin{equation}
		\hat{\mathbf{H}}_k = [\mathbf{d}_{1,k}\mathbf{g}_{1,k}^H\varphi_{1,k} \quad \mathbf{d}_{2,k}\mathbf{g}_{2,k}^H\varphi_{2,k} \quad\cdots\quad \mathbf{d}_{S,k}\mathbf{g}_{S,k}^H\varphi_{S,k}],
		\label{eq:tilde_H_k}
	\end{equation}
	where $\mathbf{d}_{s,k} = \sum_{l=1}^{L_{s,k}} a_{s,k,l}e^{-j2\pi \frac{r}{T}\tau_{s,k,l}^{\text{UT}}}\mathbf{d}_{s,k,l}$. 
	Note that $\mathbf{H}_{s,k} = \mathbf{d}_{s,k}\mathbf{g}_{s,k}^H$, which indicates that $\text{rank}(\mathbf{H}_{s,k}) = 1, \forall s$. This rank deficiency restricts the system to single-stream transmission per UT. Conversely, (\ref{eq:tilde_H_k}) shows that $\hat{\mathbf{H}}_{k}$ has a rank greater than one (i.e., $\text{rank}(\hat{\mathbf{H}}_{k}) > 1$), scaling with the number of cooperating SATs.
	These results demonstrate that through multi-SAT cooperation, multi-stream transmission to a single UT becomes feasible, enabling significant performance improvements compared to non-cooperative scenarios.
	
	In what follows, we characterize the statistical properties of $\hat{\mathbf{H}}_k$ in (\ref{eq:tilde_H_k}). 
	For SAT communications, $\hat{\mathbf{H}}_{s,k}$ can be modeled using the Rician channel model \cite{Abdi2001estimation} as:
	\begin{align}
		\hat{\mathbf{H}}_{s,k} = \mathbf{d}_{s,k}\mathbf{g}_{s,k}^H\varphi_{s,k}  =\sqrt{\frac{\kappa_{s,k}\gamma_{s,k}}{2(\kappa_{s,k}+1)}}\mathbf{H}_{s,k}^{\text{LoS}}\varphi_{s,k} + \mathbf{H}_{s,k}^{\text{NLoS}}\varphi_{s,k},
		\label{eq:Rician_H_s_k}
	\end{align}
	where $\gamma_{s,k} = \mathbb{E}\{\text{tr}(\hat{\mathbf{H}}_{s,k}\hat{\mathbf{H}}_{s,k}^{H})\}$ denotes the large-scale fading coefficient between the $s$-th SAT and the $k$-th UT, and $\kappa_{s,k}$ is the Rician K-factor.
	Here, we model $\mathbf{d}_{s,k}$ as a Rician fading vector. Specifically, we have $\mathbf{d}_{s,k} = \bar{\mathbf{d}}_{s,k} + \tilde{\mathbf{d}}_{s,k}$, where $\bar{\mathbf{d}}_{s,k} = \sqrt{\frac{\kappa_{s,k}\gamma_{s,k}}{2(\kappa_{s,k}+1)}}(1+j)\mathbf{d}_{s,k,1}$ represents the deterministic LoS component at the receiver, and $\tilde{\mathbf{d}}_{s,k}$ corresponds to the random scattering component (i.e., non-line-of-sight (NLoS)), whose elements follow an i.i.d. $\mathcal{CN}(0, \frac{\gamma_{s,k}}{N(\kappa_{s,k} + 1)})$ distribution.	
	Accordingly, we have $\mathbf{H}_{s,k}^{\text{LoS}} = (1+j)\mathbf{d}_{s,k,1}\mathbf{g}_{s,k}^{H}$ and $\mathbf{H}_{s,k}^{\text{NLoS}} = \tilde{\mathbf{d}}_{s,k}\mathbf{g}_{s,k}^H$.
	
	We define $\beta_{s,k}^{\text{LoS}} \triangleq \frac{\kappa_{s,k}\gamma_{s,k}}{2(\kappa_{s,k} + 1)}$.
	Based on the channel model in (\ref{eq:Rician_H_s_k}), the statistical mean of $\hat{\mathbf{H}}_k$, denoted as $\bar{\mathbf{H}}_k$, is given by:
	\begin{align}
		&\bar{\mathbf{H}}_k =\nonumber \\ &\Big[\sqrt{\beta_{1,k}^{\text{LoS}}}\mathbf{H}_{1,k}^{\text{LoS}}\bar{\varphi}_{1,k}\; \sqrt{\beta_{2,k}^{\text{LoS}}}\mathbf{H}_{2,k}^{\text{LoS}}\bar{\varphi}_{2,k}\; \cdots \; \sqrt{\beta_{S,k}^{\text{LoS}}}\mathbf{H}_{S,k}^{\text{LoS}}\bar{\varphi}_{S,k}\Big],
		\label{eq:mean_H_k}
	\end{align}
	where $\bar{\varphi}_{s,k} = \mathbb{E}\{\varphi_{s,k}\}$ denotes the expected phase error. Furthermore, the second-order statistic of $\hat{\mathbf{H}}_{s,k}$, $\mathbb{E}\{\hat{\mathbf{H}}_{s_1,k}^H\hat{\mathbf{H}}_{s_2, k}\}$, is given by:
	\begin{align}
		&\mathbb{E}\{\hat{\mathbf{H}}_{s_1,k}^H\hat{\mathbf{H}}_{s_2, k}\}  \nonumber \\
		&\;= \left\{ 
		\begin{array}{rcl}
			\gamma_{s,k}\mathbf{g}_{s,k}\mathbf{g}_{s,k}^H, &  s_1 = s_2 = s \\ 
			\bar{\mathbf{d}}_{s_1, k}^H\bar{\mathbf{d}}_{s_2, k}\bar{\varphi}_{s_1,k}^{*}\bar{\varphi}_{s_2,k}\mathbf{g}_{s_1,k}\mathbf{g}_{s_2,k}^H, & s_1 \neq s_2.
		\end{array}\right.
		\label{eq:E_hat_H_k}
	\end{align}
	Given (\ref{eq:mean_H_k}) and (\ref{eq:E_hat_H_k}), $\mathbb{E}\{\hat{\mathbf{H}}_k^H\hat{\mathbf{H}}_k\}$ can be expressed as:
	\begin{gather}
		\mathbb{E}\{\hat{\mathbf{H}}_k^H\hat{\mathbf{H}}_k\} = \tilde{\mathbf{H}}_k^H \tilde{\mathbf{H}}_k, \nonumber \\
		\tilde{\mathbf{H}}_k^H = [\bar{\mathbf{H}}_k^H \; \text{diagblk}\{[\sqrt{\rho_{1,k}}\mathbf{g}_{1,k},\dots, \sqrt{\rho_{S,k}}\mathbf{g}_{S,k}]\}],
		\label{eq:E_HH}
	\end{gather}
	where $\rho_{s,k} = \frac{\gamma_{s,k}(1 + \kappa_{s,k}(1 - |\bar{\varphi}_{s,k}|^2))}{(\kappa_{s,k} + 1)}$. Note that both $\bar{\mathbf{H}}_k$ and $\tilde{\mathbf{H}}_k^H\tilde{\mathbf{H}}_k$ will be used to compute the precoding matrix under the sCSI condition. 
	
	The computation of $\tilde{\mathbf{H}}_k^H$ relies on the statistical information of $\{\bar{\mathbf{H}}_k\}$, $\{\mathbf{g}_{s,k}\}$, $\{\gamma_{s,k}\}$, $\{\kappa_{s,k}\}$, and $\{\bar{\varphi}_{s,k}\}$. This statistical CSI can be obtained through UT feedback or UL channel estimation. Since statistical reciprocity exists between the UL and DL channels, each SAT can first estimate its sCSI from UL pilot signals and subsequently use it for DL transmission \cite{Ziyu2026Decoupled}. In particular, accurate directional cosines can be acquired from satellite ephemeris and global navigation satellite system (GNSS) positioning \cite{Ke-Xin2024Ergodic}. 
	
	Furthermore, the sCSI can be updated whenever the SATs receive UL pilot signals. Taking $\gamma_{s,k}$ as an example, let $\hat{\gamma}_{s,k}^t$ denote its estimate at time instant $t$, which can be computed as:
	\begin{equation}
		\hat{\gamma}_{s,k}^t = \alpha \hat{\gamma}_{s,k} + (1-\alpha)\hat{\gamma}_{s,k}^{t-1},
	\end{equation}
	where $\hat{\gamma}_{s,k}$ represents the estimate of $\gamma_{s,k}$ based on the most recently received pilot signals, and $\alpha \in [0, 1]$ is a tunable weight factor balancing the current and previous estimates. 
	It is worth noting that the update period of the sCSI can be aligned with the UL pilot signal transmission interval, which is typically $20$~ms in terrestrial communications \cite{38.214_3GPP}.  
	On the other hand, the sCSI is assumed to be quasi-static as long as the SATs and UTs are within a certain range. Therefore, the update period for the precoding matrices can exceed $20$~ms. As reported in \cite{Ziyu2026Decoupled} and shown in the simulation results, an update period of $100$~ms results in negligible performance degradation.

	\section{Locally Optimal iCSI Precoder Design}
	\label{sec:optimal_design}
	In this section, we present the iCSI precoder design subject to both TPC and PAPC, which achieves a locally optimal solution. In highly dynamic LEO SAT communications, accurate iCSI acquisition is practically challenging. Consequently, the iCSI-based design, evaluated under a favorable assumption of known instantaneous channel and impairment information, serves as a performance upper bound for the sCSI-based algorithms rather than a practically matched benchmark. Furthermore, as shown in Section \ref{sec:Optimal precoder design based on sCSI}, the algorithmic framework provided in this section applies to the sCSI-based algorithm designs.
	
	Let $\hat{\mathbf{s}}_k \in \mathbb{C}^{d_k \times 1}$ denote the received multi-stream signal at the $k$-th UT. Based on the transmission model given in (\ref{eq:y_k}), $\hat{\mathbf{s}}_k $ is given by:
	\begin{align}
		\hat{\mathbf{s}}_k = & \mathbf{F}_k^H\sum_{s=1}^{S}\mathbf{H}_{s,k}\Big(\mathbf{W}_{s,k}\mathbf{s}_k\varphi_{s,k} + \sum_{k'\neq k}^K\mathbf{W}_{s,k'}\breve{\mathbf{s}}_{s,k'}\varphi_{s,k',k}\Big)\nonumber \\ 
		& \quad + \mathbf{F}_k^H\mathbf{n}_k,
		\label{eq:hat_s_k}
	\end{align}
	where $\mathbf{F}_k\in \mathbb{C}^{N\times d_k}$ denotes the receiving matrix for $\mathbf{s}_k$. We assume that the information of different UTs is independent, $\mathbb{E}\{\mathbf{s}_k\mathbf{s}_k^H\} = \mathbf{I}_{d_k}$, and $\mathbb{E}\{\breve{\mathbf{s}}_{s,k'}\breve{\mathbf{s}}_{s, k'}^H\} = \mathbf{I}_{d_{k'}}$. Due to the independence of inter-SAT interference as described in Remark \ref{remark:interference_independent}, we approximate $\mathbb{E}\{\breve{\mathbf{s}}_{s_1,k'}\breve{\mathbf{s}}_{s_2, k'}^H\} \approx \mathbf{0}$.
	Based on the above analysis, the MSE matrix for the $k$-th UT is given by:
	\begin{align}
		\text{MSE}_k = \;& \mathbb{E}\{(\hat{\mathbf{s}}_k - \mathbf{s}_k)(\hat{\mathbf{s}}_k - \mathbf{s}_k)^H\} \nonumber \\
		= \;& \mathbf{F}_{k}^H(\hat{\mathbf{H}}_{k}\mathbf{W}_{k}\mathbf{W}_{k}^H\hat{\mathbf{H}}_k^H + \mathbf{R}'_z)\mathbf{F}_k - \mathbf{F}_k^H\hat{\mathbf{H}}_k\mathbf{W}_k \nonumber \\
		& - \mathbf{W}_k^H\hat{\mathbf{H}}_k^H\mathbf{F}_k + \mathbf{I}_{d_k},
		\label{eq:MSE}
	\end{align}
	where
	\begin{gather}
		\mathbf{R}'_{z} = \hat{\mathbf{H}}_{k}\bigg(\sum_{k'\neq k}^K \tilde{\mathbf{W}}_{k'}\tilde{\mathbf{W}}_{k'}^{H}\bigg)\hat{\mathbf{H}}_{k}^{H} +\sigma_{k}^2\mathbf{I}_N, \nonumber \\
		\mathbf{W}_{k} = [\mathbf{W}_{1,k}^{T}\; \mathbf{W}_{2,k}^{T} \;\cdots\; \mathbf{W}_{S,k}^{T}]^T \in \mathbb{C}^{SM\times d_k}, \nonumber \\
		\tilde{\mathbf{W}}_{k'} = \text{diagblk}\{[\mathbf{W}_{1,k'}, \mathbf{W}_{2,k'}, \dots,  \mathbf{W}_{S,k'}]\} \in \mathbb{C}^{SM\times Sd_{k'}}.
	\end{gather}
	The linear receiver that minimizes the trace of the MSE matrix is the well-known Wiener filter \cite{kailath2000linear}, given by:
	\begin{equation}
		\mathbf{F}_{k}^{\text{opt}} = (\hat{\mathbf{H}}_k\mathbf{W}_k\mathbf{W}_k^H\hat{\mathbf{H}}_k^H + \mathbf{R}'_z)^{-1}\hat{\mathbf{H}}_k\mathbf{W}_k.
		\label{eq:G_k_opt}
	\end{equation}
	Substituting $\mathbf{F}_k^{\text{opt}}$ into (\ref{eq:MSE}) yields the MMSE matrix for the $k$-th UT:
	\begin{equation}
		\text{MMSE}_k
		=  (\mathbf{I}_{d_k} + \mathbf{W}_k^H \hat{\mathbf{H}}_k^H (\mathbf{R}'_z)^{-1}\hat{\mathbf{H}}_k\mathbf{W}_k)^{-1}.
		\label{eq:MSE_G_k_opt}
	\end{equation}
	Furthermore, the achievable rate for the $k$-th UT, $r_k$, derived from (\ref{eq:hat_s_k}), is given by:
	\begin{equation}
		r_k = \log\det(\mathbf{I}_{d_k} + \mathbf{W}_k^H\hat{\mathbf{H}}_k^H(\mathbf{R}'_z)^{-1}\hat{\mathbf{H}}_k\mathbf{W}_k).
		\label{eq:r_k}
	\end{equation}
	
	\subsection{Total Power Constraint}
	The sum-rate maximization problem subject to the TPC is formulated as follows:
	\begin{gather}
		\max_{\{\mathbf{W}_k\}}\sum_{k=1}^{K}r_k \quad
		\text{s.t.} \; \sum_{k=1}^{K}\text{tr}\big(\mathbf{W}_k\mathbf{W}_k^H\big) \leq P,
		\label{eq:p_1_TPC} 
	\end{gather}
	where $P$ denotes the maximum available transmit power for the cooperative SATs. 
	To solve problem (\ref{eq:p_1_TPC}), we first notice that $r_k = \log\det(\text{MMSE}_k^{-1})$. We then formulate an equivalent block-convex problem for (\ref{eq:p_1_TPC}), which is given by:
	\begin{gather}
		\min_{\{\mathbf{W}_k, \mathbf{F}_k, \mathbf{\Gamma}_k\}} \sum_{k=1}^{K}\text{tr}(\mathbf{\Gamma}_k \text{MSE}_k) - \log\det(\mathbf{\Gamma}_k)\nonumber \\
		\text{s.t.} \sum_{k=1}^{K}\text{tr}(\mathbf{W}_k\mathbf{W}_k^H) \leq P,
		\label{eq:p_2_TPC}
	\end{gather}
	where $\mathbf{\Gamma}_k \in \mathbb{C}^{d_k \times d_k}$ can be interpreted as the auxiliary weight matrix for the MSE matrix of the $k$-th UT. Note that when $\mathbf{\Gamma}_k = \text{MMSE}_k^{-1}$ and $\mathbf{F}_k = \mathbf{F}_k^{\text{opt}}$, problem (\ref{eq:p_2_TPC}) is equivalent to problem (\ref{eq:p_1_TPC}).
	Such a formulation is called the WMMSE approach \cite{Shi2011WMMSE}.

	Problem (\ref{eq:p_2_TPC}) can be solved using the block coordinate descent (BCD) method. As proven in \cite{Shi2011WMMSE}, the BCD method converges to a nontrivial stationary point of (\ref{eq:p_1_TPC}). Specifically, the objective function of (\ref{eq:p_2_TPC}) is minimized by sequentially fixing two of the three variables, $\{\mathbf{W}_k\}$, $\{\mathbf{F}_k\}$, and $\{\mathbf{\Gamma}_k\}$, and updating the remaining one. Since the objective function of (\ref{eq:p_2_TPC}) is convex with respect to $\mathbf{\Gamma}_k$, we have $\mathbf{\Gamma}_k^{\text{opt}}= \text{MSE}_k^{-1}$. Furthermore, the optimal receiving matrix $\mathbf{F}_k$ is given by $\mathbf{F}_k^{\text{opt}}$ in (\ref{eq:G_k_opt}). Finally, the optimization of $\{\mathbf{W}_k\}$ can be formulated as the following problem:
	\begin{align}
		\min_{\{\mathbf{W}_k\}}\sum_{k=1}^K \text{tr}(\mathbf{\Gamma}_k\text{MSE}_k) \quad \text{s.t.} \sum_{k=1}^K \text{tr}(\mathbf{W}_k\mathbf{W}_k^H) \leq P.
		\label{eq:p_3_TPC}
	\end{align}
	When $\{\mathbf{F}_k\}$ and $\{\mathbf{\Gamma}_k\}$ are fixed, (\ref{eq:p_3_TPC}) becomes a convex quadratic problem with respect to $\{\mathbf{W}_k\}$, which can be solved using standard convex optimization algorithms. However, the Lagrange multiplier method provides a more efficient approach. Specifically, the Lagrangian of (\ref{eq:p_3_TPC}) is expressed as:
	\begin{align}
		L(\{\mathbf{W}_k\}, \mu) = & \sum_{k=1}^{K}\text{tr}\Big(\mathbf{\Gamma}_k\Big(\mathbf{F}_{k}^H(\hat{\mathbf{H}}_{k}\mathbf{W}_{k}\mathbf{W}_{k}^H\hat{\mathbf{H}}_k^H + \mathbf{R}'_z)\mathbf{F}_k \nonumber \\ 
		&-\mathbf{F}_k^H\hat{\mathbf{H}}_k\mathbf{W}_k
		- \mathbf{W}_k^H\hat{\mathbf{H}}_k^H\mathbf{F}_k + \mathbf{I}_{d_k}\Big)\Big)  \nonumber \\ 
		&+ \mu\Big(\sum_{k=1}^{K} \text{tr}(\mathbf{W}_k\mathbf{W}_k^H)- P\Big),
		\label{eq:Lagrange_TPC}
	\end{align}
	where $\mu$ denotes the Lagrange multiplier for the TPC.
	By taking the gradient of (\ref{eq:Lagrange_TPC}) with respect to $\mathbf{W}_k$ and setting it to $\mathbf{0}$, we obtain the structure of $\mathbf{W}_k$ as:
	\begin{align}
		\mathbf{W}_k(\mu) =& \Big(\hat{\mathbf{H}}_k^H\mathbf{F}_k\mathbf{\Gamma}_k\mathbf{F}_k^H\hat{\mathbf{H}}_k + \sum_{k' \neq k}\hat{\mathbf{F}}_{k'}\hat{\mathbf{\Gamma}}_{k'}\hat{\mathbf{F}}_{k'}^H + \mu \mathbf{I}_{SM}\Big)^{-1} \nonumber \\ &\quad\cdot\hat{\mathbf{H}}_k^H\mathbf{F}_k\mathbf{\Gamma}_k,
		\label{eq:W_opt_TPC}
	\end{align}
	where $\hat{\mathbf{F}}_{k'} = \text{diagblk}\{[\hat{\mathbf{H}}_{1,k'}^H\mathbf{F}_{k'},\dots, \hat{\mathbf{H}}_{S,k'}^H\mathbf{F}_{k'}]\} \in \mathbb{C}^{SM\times Sd_{k'}}$ and $\hat{\mathbf{\Gamma}}_{k'} = \text{diagblk}\{[\mathbf{\Gamma}_{k'},\dots, \mathbf{\Gamma}_{k'}]\} \in \mathbb{C}^{Sd_{k'} \times Sd_{k'}}$. The Lagrange multiplier $\mu \geq 0$ must be chosen such that the complementary slackness condition of the TPC in (\ref{eq:p_3_TPC}) is satisfied. This implies that $\sum_{k=1}^K \text{tr}(\mathbf{W}_k(\mu)\mathbf{W}_k(\mu)^H) = P$, which is equivalent to the following condition:
	\begin{equation}
		\sum_{k=1}^K\text{tr}((\mathbf{\Lambda}_k + \mu \mathbf{I}_{SM})^{-2}\mathbf{\Theta}_k) = P,
		\label{eq:mu_TPC}
	\end{equation}
	where
	\begin{equation}
		\mathbf{\Theta}_k = \mathbf{U}_k^H\hat{\mathbf{H}}_k^H\mathbf{F}_k\mathbf{\Gamma}_k\mathbf{\Gamma}_k^H\mathbf{F}_k^H\hat{\mathbf{H}}_k\mathbf{U}_k,
		\label{eq:Theta_k}
	\end{equation}
	and 
	$\mathbf{U}_k\mathbf{\Lambda}_k\mathbf{U}_k^H$ represents the eigenvalue decomposition (EVD) of $\hat{\mathbf{H}}_k^H\mathbf{F}_k\mathbf{\Gamma}_k\mathbf{F}_k^H\hat{\mathbf{H}}_k + \sum_{k' \neq k}^K\hat{\mathbf{F}}_{k'}\hat{\mathbf{\Gamma}}_{k'}\hat{\mathbf{F}}_{k'}^H$. Since $\mathbf{\Lambda}_k$ is a diagonal matrix, the left-hand side of (\ref{eq:mu_TPC}) is a monotonically decreasing function of $\mu$ for $\mu \geq 0$. Therefore, the optimal value of $\mu$, denoted as $\mu^*$, can be obtained by using a simple bisection method. The overall algorithm for solving problem (\ref{eq:p_1_TPC}) is summarized in Algorithm \ref{algm:optimal precoder design_TPC_iCSI}.
	\begin{algorithm}
		\caption{Locally Optimal Precoder Design Under TPC}\label{algm:optimal precoder design_TPC_iCSI}
		\begin{algorithmic}[1]
			\STATE Initialize the precoding matrices $\mathbf{W}_{k} = \mathbf{W}_{k}^{\text{init}}, \forall k$, set the iteration index $n = 0$, and define the maximum number of iterations $N_{\text{iter}}$.
			\STATE \textbf{Repeat}
			\STATE\quad$\mathbf{\Gamma}'_k = \mathbf{\Gamma}_k, \forall k$.
			\STATE\quad$\mathbf{F}_{k} = (\hat{\mathbf{H}}_k\mathbf{W}_k\mathbf{W}_k^H\hat{\mathbf{H}}_k^H + \mathbf{R}'_z)^{-1}\hat{\mathbf{H}}_k\mathbf{W}_k, \forall k$.
			\STATE\quad $\mathbf{\Gamma}_k = \mathbf{I}_{d_k} + \mathbf{W}_k^H\hat{\mathbf{H}}_k^H(\mathbf{R}'_z)^{-1}\hat{\mathbf{H}}_k\mathbf{W}_k, \forall k$.
			
			\STATE\quad Use the bisection method to calculate $\mu^*$ based on (\ref{eq:mu_TPC}). 
			
			\STATE\quad $\mathbf{W}_k = \big(\hat{\mathbf{H}}_k^H\mathbf{F}_k\mathbf{\Gamma}_k\mathbf{F}_k^H\hat{\mathbf{H}}_k + \sum_{k' \neq k}\hat{\mathbf{F}}_{k'}\hat{\mathbf{\Gamma}}_{k'}\hat{\mathbf{F}}_{k'}^H + \mu^{*} \mathbf{I}_{SM}\big)^{-1}\hat{\mathbf{H}}_k^H\mathbf{F}_k\mathbf{\Gamma}_k, \forall k$.
			
			\STATE\quad \textbf{If} $n \geq N_{\text{iter}}$ or $\sum_{k=1}^{K}\big|\log\det(\mathbf{\Gamma}_k) -\log\det( \mathbf{\Gamma}'_k)\big| \leq \epsilon$:
			
			\STATE\qquad Assign $\mathbf{W}_{k}^{\text{*}} = \mathbf{W}_{k}, \forall k$ and terminate the algorithm.
			
			\STATE\quad \textbf{Else}: 
			
			\STATE\qquad Set $n = n + 1$.
			
			\STATE \quad \textbf{End If}
			
			\STATE \textbf{until the termination condition is met.}
		\end{algorithmic}
	\end{algorithm}
	
	Complexity analysis: The computational complexity is dominated by the EVD operation in (\ref{eq:Theta_k}) and matrix inversion operation in line 7 of Algorithm \ref{algm:optimal precoder design_TPC_iCSI}, which scale as $\mathcal{O}(S^3M^3K)$. Given $N_i$ iterations, the total complexity is $\mathcal{O}(S^3M^3KN_i)$.
	
	\subsection{Per-antenna Power Constraint}
	Instead of the ideal TPC, a PAPC might be more practical for cooperative systems. Under the PAPC condition, the sum-rate maximization problem can also be solved using the WMMSE framework. The problem can be formulated as:
	\begin{gather}
		\min_{\{\mathbf{W}_k, \mathbf{F}_k, \mathbf{\Gamma}_k\}} \sum_{k=1}^K \text{tr}(\mathbf{\Gamma}_k\text{MSE}_k) - \log\det(\mathbf{\Gamma}_k) \nonumber \\
		\text{s.t.}\quad \sum_{k=1}^{K}[\mathbf{W}_k\mathbf{W}_k^H]_{m,m} \leq P_m, \forall m,
		\label{eq:P_1_PAPC}
	\end{gather}
	where $P_m = P/SM$, for $m = 1, \dots, SM$, denotes the maximum transmit power of the $m$-th antenna. Note that $\{\mathbf{F}_k\}$ and $\{\mathbf{\Gamma}_k\}$ can be updated similarly to the TPC case, and the problem for updating $\{\mathbf{W}_k\}$ is given by:
	\begin{equation}
		\min_{\{\mathbf{W}_k\}}\sum_{k=1}^K \text{tr}(\mathbf{\Gamma}_k\text{MSE}_k) \quad \text{s.t.} \; \sum_{k=1}^{K}[\mathbf{W}_k\mathbf{W}_k^H]_{m,m} \leq P_m, \forall m.
		\label{eq:PAPC_W_update}
	\end{equation}
	Since there are generally $SM$ power constraints in (\ref{eq:PAPC_W_update}), a simple bisection search relying on the Lagrange multiplier method is no longer applicable.

	To improve the computational efficiency of solving (\ref{eq:P_1_PAPC}), we consider employing the BCD method to sequentially optimize $\{\mathbf{\Gamma}_k\}$, $\{\mathbf{F}_k\}$, and $\{\bar{\mathbf{w}}_{k,m}\}$, where $\bar{\mathbf{w}}_{k,m} \in \mathbb{C}^{d_k \times 1}$ denotes the $m$-th column of $\mathbf{W}_{k}^H$. In other words, instead of directly optimizing $\{\mathbf{W}_k\}$, we optimize each column of $\mathbf{W}_k^H$ sequentially. Problem (\ref{eq:PAPC_W_update}) is decoupled into $SM$ subproblems, and the $m$-th subproblem is formulated as:
	\begin{gather}
		\min_{\{\bar{\mathbf{w}}_{k,m}\}_{k=1}^K}\sum_{k=1}^K\text{tr}\Big(\mathbf{\Gamma}_k\mathbf{F}_k^H\Big(\sum_{m=1}^{SM} (\hat{\mathbf{h}}_{k,m}\bar{\mathbf{w}}_{k,m}^H)\sum_{m=1}^{SM} (\bar{\mathbf{w}}_{k,m}\hat{\mathbf{h}}_{k,m}^H) \nonumber \\
		+\sum_{k'\neq k}\sum_{m=1}^{SM} (\hat{\mathbf{h}}_{k,m}\tilde{\mathbf{w}}_{k',m}^H)\sum_{m=1}^{SM} (\tilde{\mathbf{w}}_{k',m}\hat{\mathbf{h}}_{k,m}^H)\Big)\mathbf{F}_k\Big) \nonumber \\  -2\text{tr}\Big(\Re\Big(\mathbf{\Gamma}_k\mathbf{F}_k^H\sum_{m=1}^{SM}(\hat{\mathbf{h}}_{k,m}\bar{\mathbf{w}}_{k,m}^H)\Big)\Big) \nonumber \\
		\text{s.t.}  \; \sum_{k=1}^K \text{tr}(\bar{\mathbf{w}}_{k,m}\bar{\mathbf{w}}_{k,m}^H) \leq P_m,
		\label{eq:P_2_PAPC}
	\end{gather}
	where $\tilde{\mathbf{w}}_{k',m} \in \mathbb{C}^{Sd_{k'}\times 1}$ is the $m$-th column of $\tilde{\mathbf{W}}_{k'}^H$, and $\hat{\mathbf{h}}_{k,m} \in \mathbb{C}^{N\times 1}$ is the $m$-th column of $\hat{\mathbf{H}}_k$.
	Next, we show that a closed-form expression for the optimal $\bar{\mathbf{w}}_{k,m}$ exists when applying the Lagrange multiplier method.
	
	Let $\mu_m \geq 0$ denote the Lagrange multiplier associated with the power constraint of the $m$-th antenna. By formulating the Lagrangian $L(\{\bar{\mathbf{w}}_{k,m}\}, \mu_m)$, taking its gradient with respect to $\bar{\mathbf{w}}_{k,m}$, and setting it to $\mathbf{0}$, we obtain the structure of the stacked vector $\bar{\mathbf{w}}_{m} = [\bar{\mathbf{w}}_{1,m}^H \bar{\mathbf{w}}_{2,m}^H \cdots \bar{\mathbf{w}}_{K,m}^H]^H \in \mathbb{C}^{d \times 1}$ as:
	\begin{gather}
		\bar{\mathbf{w}}_{m}(\mu_m) = \frac{\mathbf{b}_{m}}{\hat{\mathbf{h}}_m^H\mathbf{\Xi}\hat{\mathbf{h}}_m + \mu_m}, 
	\end{gather}
	where $d = \sum_{k=1}^{K}d_k$ denotes the total number of signal streams transmitted to all served UTs, $\hat{\mathbf{h}}_{m} = [\hat{\mathbf{h}}_{1,m}^H,\dots, \hat{\mathbf{h}}_{K,m}^H]^H\in \mathbb{C}^{KN\times 1}$, $\mathbf{\Xi} = \text{diagblk}\{[\mathbf{\Xi}_1,\dots, \mathbf{\Xi}_K]\}\in \mathbb{C}^{KN \times KN}$ with $\mathbf{\Xi}_k = \mathbf{F}_k\mathbf{\Gamma}_k\mathbf{F}_k^H$. Furthermore, $\mathbf{b}_m$ is given by:
	\begin{align}
		\mathbf{b}_{m} =
		\mathbf{\Sigma}\hat{\mathbf{h}}_{m}
		- \Big(\sum_{m' \in \mathcal{M}_s, m'\neq m}\bar{\mathbf{w}}_{m'}\hat{\mathbf{h}}_{m'}^H
		+\sum_{m'\notin \mathcal{M}_s}
		\mathbf{\Omega}_{m'}\Big)\mathbf{\Xi}\hat{\mathbf{h}}_m,
	\end{align}
	where $\mathbf{\Sigma} = \text{diagblk}\{[\mathbf{\Gamma}_1\mathbf{F}_1^H,\dots, \mathbf{\Gamma}_K\mathbf{F}_K^H]\} \in \mathbb{C}^{d \times KN}$ and $\mathbf{\Omega}_{m'} = \text{diagblk}\{[\bar{\mathbf{w}}_{1,m'}\hat{\mathbf{h}}_{1,m'}^H, \dots, \bar{\mathbf{w}}_{K,m'}\hat{\mathbf{h}}_{K,m'}^H]\} \in \mathbb{C}^{d \times KN}$.
	Here, we assume that the considered $m$-th antenna belongs to the $s$-th SAT, where $\mathcal{M}_s$ denotes the antenna index set for this SAT (i.e., $\mathcal{M}_s = \{(s-1)M + 1,\dots, sM\}$).
	
	Similar to the TPC case, $\bar{\mathbf{w}}_{k,m}(\mu_m)$ must satisfy the complementary slackness condition for the PAPC, which is given by $\sum_{k=1}^K \text{tr}(\bar{\mathbf{w}}_{k,m}(\mu_m)\bar{\mathbf{w}}_{k,m}^H(\mu_m)) = P_m$. This yields:
	\begin{equation}
		\frac{\|\mathbf{b}_{m}\|_2^2}{(\hat{\mathbf{h}}_m^H\mathbf{\Xi}\hat{\mathbf{h}}_m + \mu_m)^2} = P_m.
	\end{equation}
	Therefore, the optimal value of $\mu_m$ is given by:
	\begin{equation}
		\mu_m^* = \max\Bigg(0, \sqrt{\frac{\|\mathbf{b}_{m}\|_2^2}{P_m}}-\hat{\mathbf{h}}_m^H\mathbf{\Xi}\hat{\mathbf{h}}_m\Bigg).
	\end{equation}
	Furthermore, $\mathbf{b}_m$ can be written as $\mathbf{b}_m = \mathbf{\Sigma}\hat{\mathbf{h}}_m - (\mathbf{C}_s - \bar{\mathbf{w}}_m\hat{\mathbf{h}}_m^H)\mathbf{\Xi}\hat{\mathbf{h}}_m$, where $\mathbf{C}_s = \sum_{m' \in \mathcal{M}_s}\bar{\mathbf{w}}_{m'}\hat{\mathbf{h}}_{m'}^H + \sum_{m'\notin \mathcal{M}_s}\mathbf{\Omega}_{m'}$. Observe that $\mathbf{C}_s$ only needs to be updated when the considered SAT changes. The overall algorithm to solve problem (\ref{eq:P_1_PAPC}) is summarized in Algorithm \ref{algm:optimal precoder design_PAPC_iCSI}.
	\begin{algorithm}
		\caption{Locally Optimal Precoder Design Under PAPC}\label{algm:optimal precoder design_PAPC_iCSI}
		\begin{algorithmic}[1]
			\STATE Initialize the precoding matrices $\mathbf{W}_{k} = \mathbf{W}_{k}^{\text{init}}, \forall k$, set the iteration index $n = 0$, and define the maximum number of iterations $N_{\text{iter}}$.
			\STATE \textbf{Repeat}
			\STATE\quad $\mathbf{\Gamma}'_k = \mathbf{\Gamma}_k, \forall k$.
			
			\STATE\quad$\mathbf{F}_{k} = (\hat{\mathbf{H}}_k\mathbf{W}_k\mathbf{W}_k^H\hat{\mathbf{H}}_k^H + \mathbf{R}'_z)^{-1}\hat{\mathbf{H}}_k\mathbf{W}_k, \forall k$.
			
			\STATE\quad $\mathbf{\Gamma}_k = \mathbf{I}_{d_k} + \mathbf{W}_k^{H}\hat{\mathbf{H}}_k^H(\mathbf{R}'_z)^{-1}\hat{\mathbf{H}}_k\mathbf{W}_k, \forall k$.
			
			\STATE\quad \textbf{For} $s = 1, \dots, S$ \textbf{do}
			
			\STATE\qquad $\mathbf{C}_s = \sum_{m' \in \mathcal{M}_s}\bar{\mathbf{w}}_{m'}\hat{\mathbf{h}}_{m'}^H + \sum_{m'\notin \mathcal{M}_s}\mathbf{\Omega}_{m'}$.
			
			\STATE\qquad \textbf{For} $m \in \mathcal{M}_s$ \textbf{do}
			
			\STATE \qquad\quad $\mathbf{b}_m = \mathbf{\Sigma}\hat{\mathbf{h}}_m - (\mathbf{C}_s - \bar{\mathbf{w}}_m\hat{\mathbf{h}}_m^H)\mathbf{\Xi}\hat{\mathbf{h}}_m$. 
			
			\STATE\qquad\quad $\mu_m^* = \max\Big(0, \sqrt{\frac{\|\mathbf{b}_{m}\|_2^2}{P_m}}-\hat{\mathbf{h}}_m^H\mathbf{\Xi}\hat{\mathbf{h}}_m\Big)$.
			
			\STATE\qquad\quad $\bar{\mathbf{w}}_{m} = \frac{\mathbf{b}_{m}}{\hat{\mathbf{h}}_m^H\mathbf{\Xi}\hat{\mathbf{h}}_m + \mu_m^*}$.
			
			\STATE\qquad \textbf{End For}
			
			\STATE\quad \textbf{End For}
			
			\STATE\quad \textbf{If} $n \geq N_{\text{iter}}$ or $\sum_{k=1}^{K}\big|\log\det(\mathbf{\Gamma}_k) -\log\det( \mathbf{\Gamma}'_k)\big| \leq \epsilon$:
			
			\STATE\qquad Assign $\mathbf{W}_{k}^{\text{*}} = \mathbf{W}_{k}, \forall k$ and terminate the algorithm.
			
			\STATE\quad \textbf{Else}: 
			
			\STATE\qquad Set $n = n + 1$ 
			
			\STATE\quad \textbf{End If}
			
			\STATE \textbf{until the termination condition is met.}
		\end{algorithmic}
	\end{algorithm}
	
	Complexity analysis: The overall computational complexity of Algorithm \ref{algm:optimal precoder design_PAPC_iCSI} is dominated by line 7, which scales as $\mathcal{O}(MdKN + (S-1)MdN)$. Because $\mathbf{C}_s$ needs to be computed $S$ times per iteration, the total computational complexity is given by $\mathcal{O}((SMdKN + S(S-1)MdN)N_i)$. Since $K$ is typically larger than $S-1$, this complexity can be simplified to $\mathcal{O}(SMdKNN_i)$.
	
	\section{Locally Optimal sCSI Precoder Design}
	\label{sec:Optimal precoder design based on sCSI}
	We now consider the locally optimal precoding design based on sCSI under the WMMSE framework.
	Note that a key step in applying the WMMSE framework to the sCSI scenario is formulating closed-form expressions for $\mathbb{E}\{r_k\}$ and $\mathbb{E}\{\text{MSE}_k\}$.
	
	Given the sCSI models in (\ref{eq:mean_H_k}) and (\ref{eq:E_HH}), we propose that $\mathbb{E}\{r_k\}$ under SAT communication can be approximated by:
	\begin{align}
		\mathbb{E}\{r_k\}& \approx \log\det\Big(\mathbf{I}_{N+S} + \tilde{\mathbf{H}}_k\mathbf{W}_k\mathbf{W}_k^H\tilde{\mathbf{H}}_k^H \nonumber \\
		&\cdot\Big(\tilde{\mathbf{H}}_k\sum_{k'\neq k}^K \big(\tilde{\mathbf{W}}_{k'}\tilde{\mathbf{W}}_{k'}^H\big)\tilde{\mathbf{H}}_k^H + \sigma_k^2 \mathbf{I}_{N+S}\Big)^{-1}\Big).
		\label{eq:E_r_k_3}
	\end{align}
	The corresponding statistical MSE, $\mathbb{E}\{\text{MSE}_k\}$, is given by:
	\begin{align}
		\mathbb{E}\{\text{MSE}_k\} = & \tilde{\mathbf{F}}_k^H(\tilde{\mathbf{H}}_k\mathbf{W}_k\mathbf{W}_k^H\tilde{\mathbf{H}}_k^H + \tilde{\mathbf{R}}'_z)\tilde{\mathbf{F}}_k - \tilde{\mathbf{F}}_k^H\tilde{\mathbf{H}}_k\mathbf{W}_k \nonumber \\& - \mathbf{W}_k^H\tilde{\mathbf{H}}_k^H\tilde{\mathbf{F}}_k + \mathbf{I}_{d_k}, \nonumber \\
		\tilde{\mathbf{R}}'_{z} =& \tilde{\mathbf{H}}_{k}\sum_{k'\neq k}^K (\tilde{\mathbf{W}}_{k'}\tilde{\mathbf{W}}_{k'}^{H})\tilde{\mathbf{H}}_{k}^{H} +\sigma_{k}^2\mathbf{I}_{N+S},
		\label{eq:MSE_sCSI}
	\end{align}
	where $\tilde{\mathbf{F}}_k$ denotes the receiving matrix for the $k$-th UT. Its optimal structure is given by:
	\begin{equation}
		\tilde{\mathbf{F}}_k^{\text{opt}} = (\tilde{\mathbf{H}}_k\mathbf{W}_k\mathbf{W}_k^H\tilde{\mathbf{H}}_k^H + \tilde{\mathbf{R}}'_z)^{-1}\tilde{\mathbf{H}}_k\mathbf{W}_k.
		\label{eq:tilde_F_opt}
	\end{equation}
	The proof of these results is provided in Appendix \ref{sec:sCSI_info}.
	
	Comparing the expressions for $\mathbb{E}\{\text{MSE}_k\}$ and $\tilde{\mathbf{F}}_k^{\text{opt}}$ with those for $\text{MSE}_k$ and $\mathbf{F}_k^{\text{opt}}$, we observe that the only difference is that the channel matrix $\hat{\mathbf{H}}_k$ is replaced by $\tilde{\mathbf{H}}_k$. 
	Therefore, the locally optimal sCSI-based precoder under the TPC condition can be computed using Algorithm \ref{algm:optimal precoder design_TPC_iCSI} by simply replacing $\hat{\mathbf{H}}_k$ with $\tilde{\mathbf{H}}_k$. Similarly, the locally optimal solution for the sCSI-based sum-rate maximization problem under the PAPC can be obtained using Algorithm \ref{algm:optimal precoder design_PAPC_iCSI} by replacing $\hat{\mathbf{H}}_k$ and $\hat{\mathbf{h}}_m$ with $\tilde{\mathbf{H}}_k$ and $\tilde{\mathbf{h}}_m$, respectively.
	
	Complexity analysis: The overall computational complexity of the locally optimal sCSI precoder design under the PAPC is also dominated by line 7 of Algorithm \ref{algm:optimal precoder design_PAPC_iCSI}. Since $\hat{\mathbf{h}}_m$ is replaced by $\tilde{\mathbf{h}}_m$, the complexity becomes $\mathcal{O}(SMdK(S+N)N_i)$. For the TPC case, the computational complexity is still dominated by the matrix inversion operation in line 7 of Algorithm \ref{algm:optimal precoder design_TPC_iCSI}, yielding a total complexity of $\mathcal{O}(S^3M^3KN_i)$.
	
	\section{Structure-Constrained Robust Precoding Design with TPC}
	\label{sec:Robust Precoding Design with TPC}
	As detailed in the complexity analysis of Algorithm \ref{algm:optimal precoder design_TPC_iCSI}, its implementation relies on an iterative process, where the complexity per iteration scales cubically with the total number of antennas (i.e., $SM$). 
	To facilitate a low-complexity solution under the TPC, this section proposes a structure-constrained precoder design. Simulation results demonstrate that the proposed approach is also robust to variations in the Rician factor and the variance of the compensated phase error.
	
	\subsection{Inter-user Interference Suppression}
	\label{sec:Inter-user interference suppression}
	To avoid the high computational complexity of iterative procedures and facilitate the receiver design, we first consider canceling the inter-UT interference at the SAT side. To achieve this, the precoder $\mathbf{W}_{s,k}$ of the $s$-th SAT for the $k$-th UT should lie in the null space of $\hat{\mathbf{H}}_{|s, k} \in \mathbb{C}^{(K-1)N\times M}$, which is defined as:
	\begin{equation}
		\hat{\mathbf{H}}_{|s,k}  \triangleq [\hat{\mathbf{H}}_{s,1}^{H}\quad \cdots \quad \hat{\mathbf{H}}_{s,k-1}^{H} \quad \hat{\mathbf{H}}_{s,k+1}^{H} \quad \cdots \quad \hat{\mathbf{H}}_{s,K}^{H}]^{H}.
		\label{eq:H_|_k}
	\end{equation} 
	Note that $\hat{\mathbf{H}}_{s,k} = \varphi_{s,k}\mathbf{d}_{s,k}\mathbf{g}_{s,k}^H$ is a rank-$1$ matrix. Therefore, computing the null space of $\hat{\mathbf{H}}_{|s,k}$ is equivalent to computing the null space of $\bar{\mathbf{G}}_{|s,k} \in \mathbb{C}^{(K-1) \times M}$, which is given by:
	\begin{equation}
		\bar{\mathbf{G}}_{|s,k} = [\mathbf{g}_{s,1}\; \cdots \mathbf{g}_{s,k-1}\; \mathbf{g}_{s,k+1} \; \cdots \mathbf{g}_{s, K}]^H.
		\label{eq:bar_G_s_k}
	\end{equation}
	An orthogonal basis for the null space of $\bar{\mathbf{G}}_{|s,k}$ can be obtained by performing a singular value decomposition (SVD) on $\bar{\mathbf{G}}_{|s,k}$. The right singular vectors corresponding to the zero singular values span this null space. We denote the matrix containing these basis vectors as $\mathbf{V}_{s,k} \in \mathbb{C}^{M\times I_{s,k}}$, where $I_{s,k} = M - \text{rank}(\bar{\mathbf{G}}_{|s,k})$. The precoder for the $k$-th UT can then be written as:
	\begin{equation}
		\mathbf{W}_k = \mathbf{V}_k\mathbf{A}_k, \quad \mathbf{V}_k = \text{diagblk}\{[\mathbf{V}_{1,k}, \mathbf{V}_{2,k}, \dots, \mathbf{V}_{S,k}]\},
		\label{eq:W_two_layer}
	\end{equation}
	where $\mathbf{A}_k \in \mathbb{C}^{I_k \times d_k}$ is a linear combination matrix with $I_k = \sum_{s=1}^S I_{s,k}$. 
	When the precoding matrix $\mathbf{W}_k$ adopts the constrained structure in (\ref{eq:W_two_layer}), the interference terms become:
	\begin{align}
		\hat{\mathbf{H}}_{k}\sum_{k'\neq k}^K \tilde{\mathbf{W}}_{k'}\tilde{\mathbf{W}}_{k'}^{H}\hat{\mathbf{H}}_{k}^{H} = \mathbf{0}_N \text{ and } \mathbf{R}'_z = \sigma_k^2 \mathbf{I}_N. 
	\end{align}
	
	\subsection{Precoding based on Statistical CSI}
	\label{sec:Precoding based on Statistical CSI}
	In this section, we first consider the problem of minimizing the sum of the traces of $\mathbb{E}\{\text{MMSE}_k\}$ across all UTs, which is formulated as:
	\begin{gather}
		\min_{\{\mathbf{W}_k\}}  \; \sum_{k=1}^{K}\text{tr}(\mathbb{E}\{\text{MMSE}_k\}) \quad
		\text{s.t.} \; \sum_{k=1}^{K}\text{tr}\big(\mathbf{W}_k\mathbf{W}_k^H\big) \leq P.
		\label{eq:MSE_optimization_total_power}
	\end{gather}
	Because inter-UT interference was suppressed in the previous subsection via the imposed precoder structure, $\mathbb{E}\{\text{MMSE}_k\}$ can be approximated based on (\ref{eq:E_MMSE_k}) as:
	\begin{equation}	
		\mathbb{E}\{\text{MMSE}_k\} \approx \mathbf{E}_k = (\mathbf{I}_{d_k} + \mathbf{W}_k^H \tilde{\mathbf{H}}_k^H\tilde{\mathbf{H}}_k\mathbf{W}_k/\sigma_k^2)^{-1}.
		\label{eq:MSE_lower_bound}
	\end{equation}
	By substituting (\ref{eq:MSE_lower_bound}) and the precoder structure $\mathbf{W}_k = \mathbf{V}_k\mathbf{A}_k$ into (\ref{eq:MSE_optimization_total_power}), we can rewrite the optimization problem as:
	\begin{gather}
		\min_{\{\mathbf{A}_k\}}   \; \sum_{k=1}^{K}\text{tr}(\mathbf{E}_k) \quad
		\text{s.t.} \; \sum_{k=1}^{K}\text{tr}\big(\mathbf{V}_k\mathbf{A}_k\mathbf{A}_k^H\mathbf{V}_k^H\big) \leq P.
		\label{eq:MSE_optimization_2}
	\end{gather}
	
	To solve (\ref{eq:MSE_optimization_2}), we first examine the feasible set of $\{\mathbf{A}_k\}$ satisfying its power constraint. Let $\mathbf{\Psi}_k = \mathbf{V}_k^H\tilde{\mathbf{H}}_k^H\tilde{\mathbf{H}}_k\mathbf{V}_k/\sigma_k^2$. According to Lemma \ref{lemma:d_lambda} on majorization theory outlined in Appendix A, for a given Hermitian matrix $\mathbf{R}$ (which, in our context, is $(\mathbf{I}_{d_k} + \mathbf{A}_k^H\mathbf{\Psi}_k\mathbf{A}_k)^{-1}$, or equivalently, $\mathbf{A}_k^H\mathbf{\Psi}_k\mathbf{A}_k$), the vector containing its diagonal elements, $\text{diag}\{\mathbf{R}\}$, is majorized by the vector $\bm{\lambda}$, which is formed by its eigenvalues. Building upon this, we establish the following theorem:
	\begin{theorem}
		Let $\mathbf{\Psi}\in\mathbb{C}^{n\times n}$ be positive semi-definite with eigenvalue decomposition (EVD) $\mathbf{\Psi} = \mathbf{U}_2\mathbf{\Lambda}_2\mathbf{U}_2^H$, and $\mathbf{A}\in\mathbb{C}^{n\times s}$ with $n \geq s$ and $\text{tr}(\mathbf{A}\mathbf{A}^H)\leq P$. Further, let the EVD lead to $\mathbf{A}^H\mathbf{\Psi}\mathbf{A}=\mathbf{U}_1\mathbf{\Lambda}_1\mathbf{U}_1^H$, and $\tilde{\mathbf{A}}=\mathbf{U}_2\mathbf{P}^{1/2}$, where the real-valued $n\times s$ matrix $\mathbf{P} = \left[ \begin{array}{c} \text{diag}(\tilde{\mathbf{p}}) \\ \mathbf{0}_{(n-s)\times s} \end{array} \right]$ with $[\tilde{\mathbf{p}}]_i = [\mathbf{\Lambda}_1]_{i,i}/[\mathbf{\Lambda}_2]_{i,i}$, $i=1,\dots,s$. Then
		\begin{gather*}
			\tilde{\mathbf{A}}^H\mathbf{\Psi}\tilde{\mathbf{A}} = \mathbf{\Lambda}_1 \quad \text{and} \quad \text{tr}(\tilde{\mathbf{A}}\tilde{\mathbf{A}}^H)\leq \text{tr}(\mathbf{A}\mathbf{A}^H).
		\end{gather*}
		\label{theorem:eigenvalues}
	\end{theorem}
	\vspace{-0.5cm}
	The proof of Theorem \ref{theorem:eigenvalues} is provided in Appendix \ref{appdix:Theorem1proof}. Theorem \ref{theorem:eigenvalues} establishes that for any feasible matrix $\mathbf{A}$, we can always find another feasible matrix $\tilde{\mathbf{A}}$ achieving the same objective value (i.e., $\text{tr}(\mathbf{A}^H\mathbf{\Psi}_k\mathbf{A}) = \text{tr}(\tilde{\mathbf{A}}^H\mathbf{\Psi}_k\tilde{\mathbf{A}})$). Moreover, the total power consumed by $\tilde{\mathbf{A}}$ is less than or equal to that of $\mathbf{A}$. Consequently, we can restrict the feasible set from $\{\mathbf{A}_k\}$ to $\{\tilde{\mathbf{A}}_k\}$ without incurring any performance loss. Since each matrix within $\{\tilde{\mathbf{A}}_k\}$ diagonalizes $\tilde{\mathbf{A}}_k^H\mathbf{\Psi}_k\tilde{\mathbf{A}}_k$, the optimization problem is effectively transformed into optimizing the diagonal variables $\mathbf{P}_k = \text{diag}\{[p_{k,1},\dots,p_{k,d_k}]\}$. 
	
	We apply Theorem \ref{theorem:eigenvalues} to problem (\ref{eq:MSE_optimization_2}) by
	setting $\mathbf{A}_k = \mathbf{U}_k\mathbf{P}_k^{1/2}$, where $\mathbf{U}_k$ consists of the eigenvectors corresponding to the $d_k$ largest eigenvalues of $\mathbf{\Psi}_k$. Consequently, problem (\ref{eq:MSE_optimization_2}) simplifies to:
	\begin{gather}
		\min_{\{p_{k,i}\}}   \sum_{k=1}^{K}\sum_{i=1}^{d_k}\big(1 + p_{k,i}\lambda_{k,i}\big)^{-1} \nonumber \\
		\text{s.t.} \sum_{k=1}^K\sum_{i=1}^{d_k}p_{k,i} \leq P,\;
		p_{k,i} \geq 0,\;
		k = 1,\dots,K,\; i = 1,\dots,d_k, 
		\label{eq:MSE_optimization_3}
	\end{gather}
	where $\lambda_{k,i}$ denotes the $i$-th eigenvalue of $\mathbf{\Psi}_k$. Since both the objective function and the constraints of problem (\ref{eq:MSE_optimization_3}) are convex, this problem can be efficiently solved using standard convex optimization tools. Notably, the transformation from $\mathbf{A}_k$ to $\tilde{\mathbf{A}}_k$ not only recasts (\ref{eq:MSE_optimization_total_power}) as the convex optimization problem in (\ref{eq:MSE_optimization_3}) but also significantly reduces the problem's dimensionality (i.e., from $I_k\times d_k$ complex variables to $d_k$ real variables), which effectively decreases the computational complexity. In fact, by applying the Lagrange multiplier method, the optimal power allocation $p_{k,i}^*$ for (\ref{eq:MSE_optimization_3}) is given by $p_{k,i}^* = \max\Big(0, \frac{1}{\sqrt{\mu^*\lambda_{k,i}}} - \frac{1}{\lambda_{k,i}}\Big)$. The optimal Lagrange multiplier, $\mu^*$, can be found via the bisection method such that the TPC $\sum_{k,i}p^*_{k,i} = P$ is satisfied. 
	
	\subsection{Sum Rate Maximization with TPC}
	We propose an approximate robust sum-rate-oriented design in this subsection. The sum-rate maximization problem subject to the TPC under the sCSI condition is formulated as:
	\begin{gather}
		\max_{\{\mathbf{W}_k\}}\sum_{k=1}^K\sum_{i=1}^{d_k}\mathbb{E}\{\log_2(1 + \text{SINR}_{k,i}^{\text{MMSE}})\} \nonumber \\
		\text{s.t.} \quad \sum_{k=1}^K \text{tr}\big(\mathbf{W}_k\mathbf{W}_k^H\big) \leq P,
		\label{eq:sum_rate_maximization_stats}
	\end{gather}
	where $\text{SINR}_{k,i}^{\text{MMSE}}$ is given by:
	\begin{gather}
		\text{SINR}_{k,i}^{\text{MMSE}} = \frac{\xi_{k,i}}{1 - \xi_{k,i}}, \quad\xi_{k,i} = \mathbf{w}_{k,i}^H\hat{\mathbf{H}}_k^H\mathbf{\Phi}_k^{-1}\hat{\mathbf{H}}_k\mathbf{w}_{k,i}, \nonumber \\
		\mathbf{\Phi}_k = \hat{\mathbf{H}}_k\sum_{k'=1}^{K}\sum_{i'=1}^{d_{k'}}(\mathbf{w}_{k',i'}\mathbf{w}_{k',i'}^H)\hat{\mathbf{H}}_k^H + \sigma_k^2 \mathbf{I}_N.
	\end{gather}
	Let $[\text{MMSE}_k]_{i,i}$ denote the $i$-th diagonal element of $\text{MMSE}_k$. By applying the Woodbury matrix identity, it can be readily shown that this diagonal element satisfies $[\text{MMSE}_k]_{i,i} = (1 + \text{SINR}_{k,i}^{\text{MMSE}})^{-1}$. 
	Consequently, by applying Jensen's inequality for bounding, problem (\ref{eq:sum_rate_maximization_stats}) can be approximately formulated as the following form:
	\begin{gather}                                     
		\min_{\{\mathbf{A}_k\}}-\prod_{k=1}^K\prod_{i=1}^{d_k}\frac{1}{[\mathbf{E}_k]_{i,i}}\quad
		\text{s.t. } \sum_{k=1}^{K}\text{tr}\big(\mathbf{V}_k\mathbf{A}_k\mathbf{A}_k^H\mathbf{V}_k^H\big) \leq P.
		\label{eq:sum_rate_maximization_stats_3}
	\end{gather}
	The detailed derivation is provided in Appendix \ref{appdix:Max_SE_problem_conversion}.
	
	Next, we demonstrate that the optimal structure of $\mathbf{A}_k$ derived in Theorem \ref{theorem:eigenvalues} is also applicable to the sum-rate maximization problem in (\ref{eq:sum_rate_maximization_stats_3}). 
	First, observe that the objective function in (\ref{eq:sum_rate_maximization_stats_3}) takes the form of $f_0(\mathbf{x}) = -\prod_{i}x_i^{-1}$, which is a Schur-concave function (the proof is provided in Appendix \ref{Appdix:Schur-Concave}).
	Because $f_0(\mathbf{x})$ is Schur-concave, the properties of majorization dictate that the vector of diagonal elements $\text{diag}\{\mathbf{E}_k\}$ is majorized by the vector of its eigenvalues $\bm{\lambda}(\mathbf{E}_k)$. This implies that:
	\begin{equation}
		-\prod_{k=1}^K\prod_{i=1}^{d_k}\frac{1}{[\mathbf{E}_k]_{i,i}} \geq -\prod_{k=1}^K\prod_{i=1}^{d_k}\frac{1}{[\bm{\lambda}(\mathbf{E}_k)]_i}.
	\end{equation}
	In Theorem \ref{theorem:eigenvalues}, we proved that for any feasible matrix $\mathbf{A}_k$ that constructs $\mathbf{E}_k$ and satisfies the power constraint in (\ref{eq:sum_rate_maximization_stats_3}), we can always find another matrix $\tilde{\mathbf{A}}_k = \mathbf{U}_k\mathbf{P}^{1/2}$. This matrix $\tilde{\mathbf{A}}_k$ makes $\mathbf{E}_k = \text{diag}(\bm{\lambda}(\mathbf{E}_k))$ and also satisfies the power constraint in (\ref{eq:sum_rate_maximization_stats_3}). 
	Thus, problem (\ref{eq:sum_rate_maximization_stats_3}) can be equivalently recast as:
	\begin{gather}
		\max_{\{p_{k,i}\}} \sum_{k=1}^K\sum_{i=1}^{d_k}\log(1+p_{k,i}\lambda_{k,i}) \nonumber \\
		\text{s.t.}\; \sum_{k=1}^K\sum_{i = 1}^{d_k} p_{k,i} \leq P,\;\; p_{k,i} \geq 0, k = 1,\dots, K, i = 1,\dots, d_k.
		\label{eq:sum_rate_maximization_stats_4}
	\end{gather}
	Since $\lambda_{k,i} \geq 0$ and $p_{k,i} \geq 0, \forall k, \forall i$, the Hessian matrix of the objective function in (\ref{eq:sum_rate_maximization_stats_4}) is negative semi-definite. Thus, the objective function is concave \cite{Boyd_Vandenberghe_2004}, and (\ref{eq:sum_rate_maximization_stats_4}) can be efficiently solved using standard convex optimization tools. 
	Similar to the MMSE criterion, by applying the Lagrange multiplier method,
	the optimal power allocation $p_{k,i}^*$ for (\ref{eq:sum_rate_maximization_stats_4}) is given by $p_{k,i}^* = \max\Big(0, \frac{1}{\mu^*} - \frac{1}{\lambda_{k,i}}\Big)$. 
	The overall algorithm for solving the sCSI-based sum-rate maximization subject to the TPC is summarized in Algorithm \ref{algm:sum_rate_max_TPC_sCSI}.
	\begin{algorithm}
		\caption{Robust Precoder Design under TPC}
		\label{algm:sum_rate_max_TPC_sCSI}
		\begin{algorithmic}[1]
			\STATE Calculate the null space matrix $\mathbf{V}_k$ for each UT by performing SVD on $\{\bar{\mathbf{G}}_{|s,k}\}, \forall s, \forall k$.
			\STATE Perform an EVD on $\mathbf{\Psi}_k = \mathbf{V}_k^H\tilde{\mathbf{H}}_k^H\tilde{\mathbf{H}}_k\mathbf{V}_k/\sigma_k^2$ to obtain the largest $d_k$ eigenvalues $\{\lambda_{k,i}\}_{i=1}^{d_k}, \forall k$ and their corresponding eigenvectors $\mathbf{U}_k \in \mathbb{C}^{I_k\times d_k}$.
			\STATE Use the bisection method to calculate $\mu^*$ such that $\sum_{k=1}^K\sum_{i=1}^{d_k}\max\Big(0, \frac{1}{\mu^*} - \frac{1}{\lambda_{k,i}}\Big) = P$.
			\STATE Calculate the power allocation $p_{k,i}^*$ as $p_{k,i}^* = \max\Big(0, \frac{1}{\mu^*} - \frac{1}{\lambda_{k,i}}\Big)$.
			\STATE Calculate the precoder as $\mathbf{W}_k^{\text{*}} = \mathbf{V}_k\mathbf{U}_k\text{diag}\{[\sqrt{p_{k,1}^*},\dots, \sqrt{p_{k,d_k}^*}]\}$.
		\end{algorithmic}
	\end{algorithm}
	
	Complexity analysis: The overall computational complexity is dominated by line 2 of Algorithm \ref{algm:sum_rate_max_TPC_sCSI}. 
	Specifically, the complexity of performing the EVD on $\{\mathbf{\Psi}_k\}, \forall k$ scales as $\mathcal{O}(I_k^3K)$. 
	
	\subsection{Low-Complexity Operation on EVD}
	The application of Theorem \ref{theorem:eigenvalues} requires performing an EVD on $\mathbf{\Psi}_k \in \mathbb{C}^{I_k \times I_k}$, which incurs a high computational complexity of $\mathcal{O}(I_k^3)$. To tackle this problem, we note that the number of transmitted data streams, $d_k$, satisfies $d_k \leq \min(S, N) \ll I_k$. Thus, we only require the $d_k$ largest  eigenvalues of  $\mathbf{\Psi}_k$ and their corresponding eigenvectors, which can be efficiently calculated using the Lanczos algorithm \cite{lanczos1950iteration}. 
	The detailed steps are outlined in Algorithm \ref{algm:Lanczos}. 
	\begin{algorithm}
		\caption{Lanczos Algorithm}
		\label{algm:Lanczos}
		\begin{algorithmic}[1]
			\STATE Initialize a random normalized vector $\mathbf{c} \in \mathbb{C}^{I_k \times 1}$. Initialize a zero matrix $\mathbf{C} \in \mathbb{C}^{I_k\times \max(S, N)}$ and set its first column: $[\mathbf{C}]_{1} = \mathbf{c}$. 
			
			\STATE \textbf{For} $i = 1,\dots, \max(S, N)$:
			
			\STATE\quad Compute $\mathbf{w} = \mathbf{\Psi}_k[\mathbf{C}]_{i}$.
			
			\STATE\quad \textbf{If} $i > 1$, update $\mathbf{w} = \mathbf{w} - \beta_{i-1}[\mathbf{C}]_{i-1}$, \textbf{End If}
			
			\STATE\quad Compute $\alpha_i = [\mathbf{C}]_i^H \mathbf{w}$. Update $\mathbf{w} = \mathbf{w} - \alpha_i [\mathbf{C}]_i$.
			
			\STATE\quad \textbf{If} $i < \max(S, N)$: 
			\STATE\qquad Compute $\beta_i = \|\mathbf{w}\|$. 
			\STATE \qquad \textbf{If} $\beta_i = 0$: 
			\STATE \qquad \quad Truncate $\mathbf{C}$ to its first $i$ columns: $\mathbf{C} = [\mathbf{C}]_{1:i}$. 
			\STATE \qquad \quad Construct the $i\times i$ tridiagonal matrix $\mathbf{T}$ as: 
			\begin{gather}
				[\mathbf{T}]_{j,j} = \alpha_j, j = 1, \dots,i, \nonumber \\
				[\mathbf{T}]_{j,j+1} =  [\mathbf{T}]_{j+1,j} = \beta_j, j = 1, \dots, i-1.
				\label{eq:T_matrix}
			\end{gather}
			\STATE \qquad\quad Terminate the iteration. 
			\STATE \qquad \textbf{Else If} $\beta_i \neq 0$: 
			\STATE \qquad \quad Set the next column of $\mathbf{C}$ as $[\mathbf{C}]_{i+1} = \frac{\mathbf{w}}{\beta_{i}}$.
			\STATE \qquad \textbf{End If}
			\STATE \quad \textbf{End If}
			\STATE \textbf{End For}

			\STATE \textbf{If} $\beta_{\max(S, N)-1} \neq 0$, construct the $\max(S, N) \times \max(S, N)$ tridiagonal matrix $\mathbf{T}$ as in (\ref{eq:T_matrix}), \textbf{End If}.
			
			\STATE Perform an EVD operation on $\mathbf{T}$ to obtain its eigenvalues $\{\lambda_{k,i}\}$ and eigenvectors $\tilde{\mathbf{U}}_k$. Let $\mathbf{U}_k = \mathbf{C}\tilde{\mathbf{U}}_k$.
			
			\STATE Select eigenvalues and eigenvectors satisfying the condition: $\| \mathbf{\Psi}_k[\mathbf{U}_k]_i - [\mathbf{U}_k]_i\lambda_{k,i}\| \leq \epsilon, \forall i$.
		\end{algorithmic}
	\end{algorithm}
	To avoid the issue of repeated eigenvalues when implementing the Lanczos algorithm, we first calculate $\max(S, N)$ eigenvalues for each UT and then remove repeated eigenvalues using line 19 of Algorithm \ref{algm:Lanczos}.
	
	The approximated eigenvalues $\{\lambda_{k,i}\}$ obtained from Algorithm \ref{algm:Lanczos} are substituted into (\ref{eq:MSE_optimization_3}) to further calculate the power control coefficients.
	Note that the overall computational complexity of Algorithm \ref{algm:Lanczos} is dominated by the $\max(S, N)$ iterations of line 3 and line 19, each of which scales as $\mathcal{O}(I_k^2\max(S, N))$. Consequently, the total computational complexity for $K$ UTs is $\mathcal{O}(I_k^2K\max(S, N))$. 
	Since $\max(S, N)\ll I_k$, the computational complexity of applying Algorithm \ref{algm:Lanczos} is significantly reduced compared to that of the standard EVD operation. It is worth mentioning that the complexity of calculating the null space matrix $\{\mathbf{V}_k\}$ (i.e., line 1 of Algorithm \ref{algm:sum_rate_max_TPC_sCSI}) is $\mathcal{O}((K-1)^2MSK)$, which is generally comparable to the complexity of the Lanczos algorithm. 
	
	\section{Simulation Results}
	\label{sec:simulations}
	We adopt the LEO Starlink SAT constellation to serve UTs uniformly distributed within a coverage area between longitudes $105^\circ$ and $120^\circ$, and latitudes $35^\circ$ and $50^\circ$. We simulate a one-hour service window comprising 60 snapshots. In each snapshot, we select SATs with elevation angles between $10^\circ$ and $90^\circ$ for all UTs. Qualified SATs are grouped into clusters of $S$ SATs. For the subsequent evaluations, 200 SAT clusters are randomly selected for the simulation. 
	The SAT communication channels are generated according to the 3GPP specifications \cite{38.811_3GPP} and the detailed parameters are summarized in Table \ref{tabl:simu_para}. 
	\begin{table}[htbp]
		\caption{Simulation Parameters}
		\begin{center}
			\begin{tabular}{c c}
				\hline
				\textbf{Parameters} & \textbf{Values} \\
				\hline
				Satellite constellation & Starlink \\
				Orbit altitudes $h_0$ & 535 $\sim$ 565 km \\
				UT distribution & Uniform \\
				UT longitudes & $105^{\circ} \sim 120^{\circ}$ \\
				UT latitudes & $35^{\circ}\sim 50^{\circ}$ \\
				Inter-SAT distance (Intra-plane) & $\sim 2000$ km \\
				Inter-SAT distance (Inter-plane) & $\sim 600$ km\\
				Central frequency $f_c$ & 2.19 GHz \cite{38.811_3GPP}\\
				Bandwidth $B$ & 20 MHz \cite{38.811_3GPP} \\
				Ionospheric loss &  1 dB~\cite{38.811_3GPP} \\
				Number of antennas per UT $N_{x'}$, $N_{y'}$ & 4, 2 \\
				Antenna spacing $d_x$, $d_y$, $d_{x'}$, $d_{y'}$ & $\lambda$, $\lambda$, $\frac{\lambda}{2}$, $\frac{\lambda}{2}$ \cite{Ke-Xin2022Downlink}\\ 
				Per-antenna gain $G_{\text{SAT}}$, $G_{\text{UT}}$ & 6 dBi, 0 dBi \cite{Ke-Xin2022Downlink}\\
				Elevation angle & $10^{\circ} \sim 90^{\circ}$\\
				\hline	 
			\end{tabular}
			\label{tabl:simu_para}
		\end{center}
	\end{table}
	
	The number of transmitted data streams significantly impacts system performance and is typically determined at the UT side. We simulate this process by solving the following problem for each UT:
	\begin{gather}
		\max_{\{p_{k,i}\}} \prod_{i=1}^{\tilde{d}_k}(1 + p_{k,i}\lambda_{k,i}) \nonumber \\
		\text{s.t.}\; \sum_{i = 1}^{\tilde{d}_k}p_{k,i} \leq \frac{P}{K}, \quad p_{k,i} \geq 0, \quad i = 1,\dots,\tilde{d}_{k}.
		\label{eq:d_k_determination}
	\end{gather}
	The above problem maximizes the achievable rate per UT under an equal power allocation assumption (i.e., $\frac{P}{K}$ per UT). We iterate $\tilde{d}_k$ from $1$ to $d_k^{\max}$ to find the optimal number of streams, $d_k^{\text{opt}}$, for rate maximization, where $d_k^{\max} = \min(S, N)$. Additionally, the distribution of the phase error $\varphi_{s,k}$ depends on the specific estimation method and is beyond the scope of this paper. Without loss of generality, we adopt the modeling approach in \cite{Xin2024Asyncrhonus, Yafei2026Statistical}. Specifically, $\varphi_{s,k} = e^{j\psi_{s,k}}$, with $\psi_{s,k} \sim \mathcal{N}(0, \varrho_{s,k}^2)$.
	
\begin{figure}[htbp]
		\centerline{\includegraphics[scale=0.43]{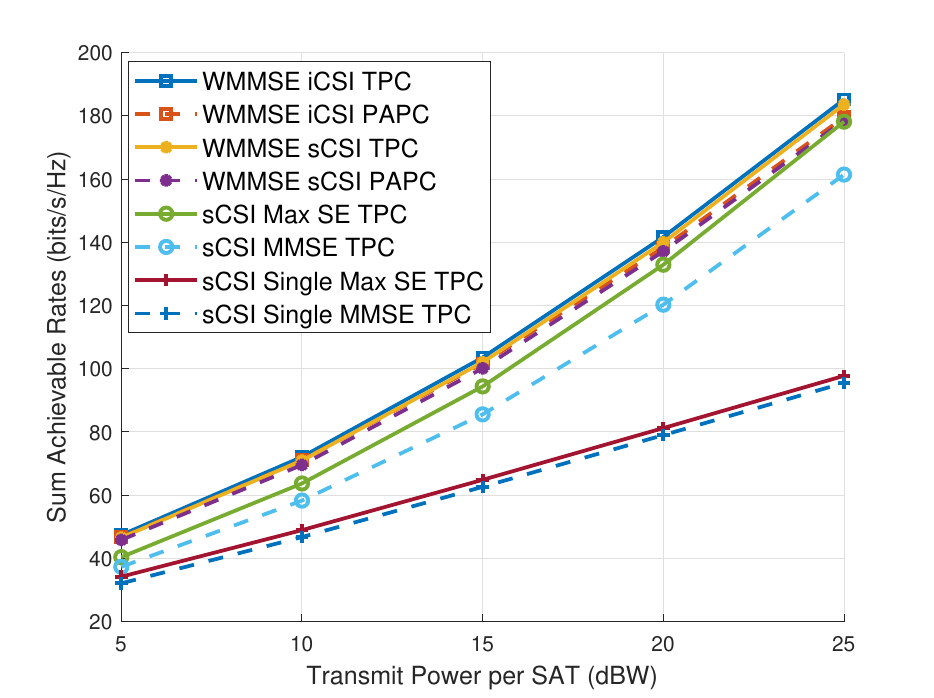}}
		\caption{Performance comparison under the total power constraint at different transmit power levels. $S = 4$; $M_x = 6, M_y = 6$; $K = 10$; $\kappa_{s,k} = 10$~dB; $\varrho_{s,k}^2 = 0.05$. (Note: since the iCSI performance in the current and following figures is obtained under the assumption of known channel and impairment information, it serves as a performance upper bound rather than a practically matched benchmark for the sCSI designs.)}
		\label{Fig:sCSI_iCSI_Comparison}
	\end{figure}

	Fig. \ref{Fig:sCSI_iCSI_Comparison} compares the performance of different precoding algorithms under both TPC and PAPC conditions. 
	The meaning of each label in the simulated results is defined as follows:
	\begin{itemize}
		\item \textbf{WMMSE} iCSI TPC/PAPC: Locally optimal sum SE maximization with iCSI (i.e., $\{\hat{\mathbf{H}}_k\}$) using the WMMSE framework, implemented via Algorithms \ref{algm:optimal precoder design_TPC_iCSI} and \ref{algm:optimal precoder design_PAPC_iCSI}, respectively.
		
		\item \textbf{WMMSE} sCSI TPC/PAPC: Locally optimal sum SE maximization with sCSI (i.e., $\{\tilde{\mathbf{H}}_k\}$) using the WMMSE framework, implemented via Algorithms \ref{algm:optimal precoder design_TPC_iCSI} and \ref{algm:optimal precoder design_PAPC_iCSI}, respectively.
		
		\item \textbf{sCSI Max SE/MMSE TPC:} Precoding design under TPC with sCSI using Algorithm \ref{algm:sum_rate_max_TPC_sCSI}, based on the achievable sum SE maximization and MMSE criteria, respectively.
		
		\item \textbf{sCSI Single Max SE/MMSE TPC:} Precoding design under TPC with sCSI using Algorithm \ref{algm:sum_rate_max_TPC_sCSI}, but strictly limited to a single stream per UT.
	\end{itemize}
	
	As shown in Fig. \ref{Fig:sCSI_iCSI_Comparison}, multi-stream algorithms significantly outperform single-stream algorithms, especially in the high transmit power regime. This is expected, as higher transmit power allows more power to be allocated to data streams with lower channel gains. We also notice that the performance gap between the ``WMMSE iCSI" upper bound and ``WMMSE sCSI" algorithms is very small, which suggests that the proposed WMMSE sCSI framework has a high approximation accuracy of the channel statistics given a small phase error variance ($\varrho_{s,k}^2 = 0.05$) and a typical Rician factor value ($\kappa_{s,k} = 10$~dB) under LEO SAT communications. Furthermore, the ``Max SE" criterion consistently outperforms the ``MMSE" criterion across the entire considered transmit power range, validating the effectiveness of the proposed approximation under the Max SE objective for rate maximization.
	
\begin{figure}[htbp]
		\centerline{\includegraphics[scale=0.3]{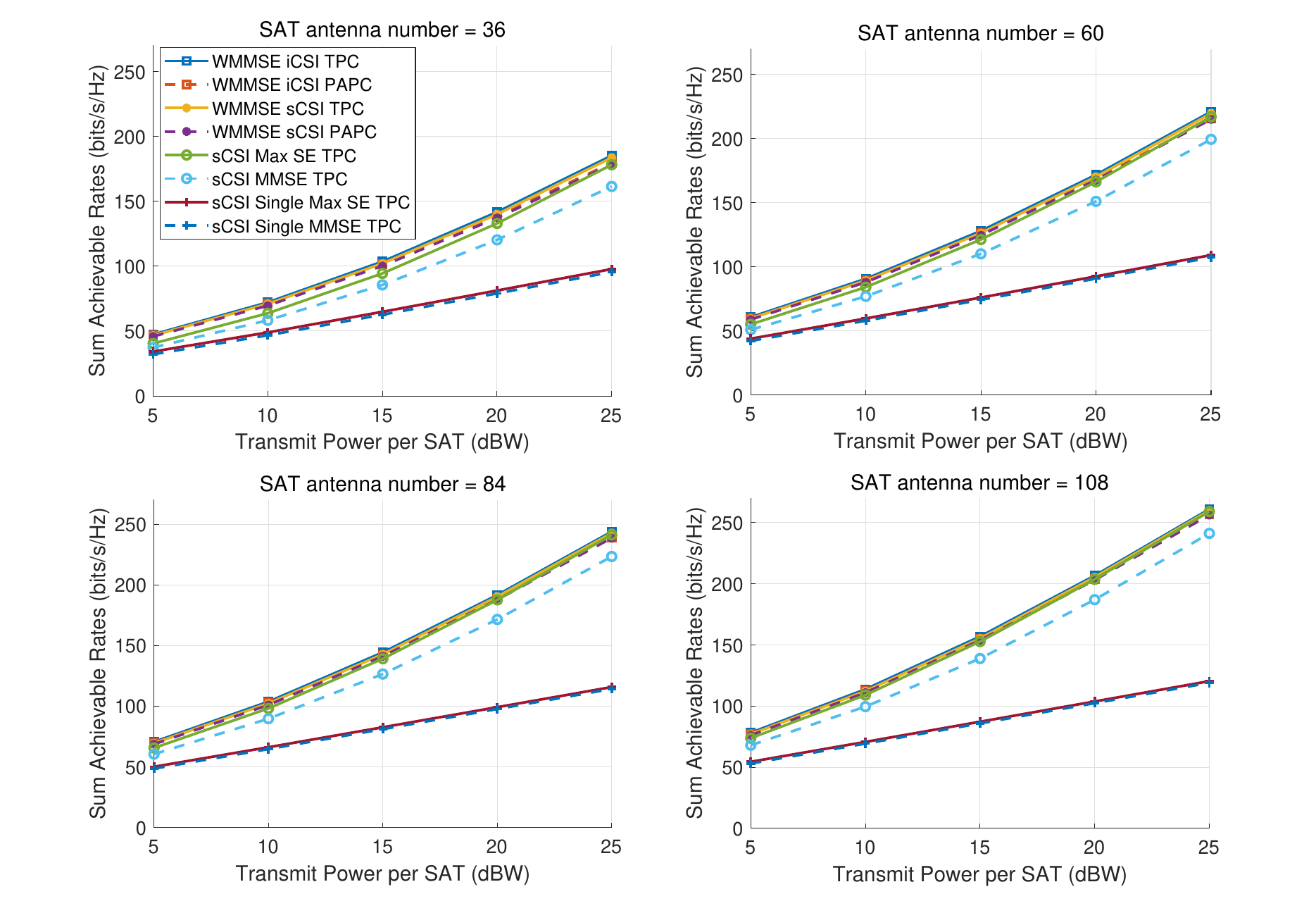}}
		\caption{Performance comparison for different numbers of antennas per SAT. $S = 4$; $K = 10$; $\kappa_{s,k} = 10$~dB; $\varrho_{s,k}^2 = 0.05$.}
		\label{Fig:Diff_Ant_Num}
	\end{figure}

	While the performance of ``sCSI Max SE TPC" approaches that of the WMMSE algorithms in Fig. \ref{Fig:sCSI_iCSI_Comparison}, a noticeable gap persists.
	This is expected, given that in Fig. \ref{Fig:sCSI_iCSI_Comparison}, the total number of transmit antennas for the cooperative SATs is 144, while the total number of antennas for all served UTs is 80.
	The null-space calculation not only suppresses inter-UT interference but also limits the dimensions of the desired signal space.
	As illustrated in Fig. \ref{Fig:Diff_Ant_Num}, increasing the number of antennas per cooperative SAT mitigates this performance gap.
	Table \ref{tab:ABC} details the percentage performance gap between ``WMMSE sCSI TPC" and ``sCSI Max SE TPC", calculated as:
	\begin{align}
		&\text{Gap} = \frac{\text{WMMSE sCSI TPC} - \text{sCSI Max SE TPC}}{\text{sCSI Max SE TPC}} \times 100\%.
		\label{eq:performance_gap_calculation}
	\end{align}
	\begin{table}[htbp]
		\centering 
		\caption{Performance gap between ``WMMSE sCSI TPC" and ``sCSI Max SE TPC" at different numbers of antennas per SAT} 
		\label{tab:ABC} 
		\begin{tabular}{|M{1.9cm}|M{0.8cm}|M{0.9cm}|M{0.9cm}|M{0.9cm}|M{0.9cm}|}
			\hline
			Transmit power per SAT & 5dBW  & 10dBW  & 15dBW  & 20dBW  & 25dBW  \\ \hline
			
			Ant Num = $36$ & 15.24\%	& 11.37\% &	8.09\% &	5.31\% &	3.05\% \\
			\hline
			
			Ant Num = $60$ & 8.23\% & 6.15\% & 4.15\% &	2.40\%	& 1.19\% \\
			\hline
			
			Ant Num = $84$ & 5.84\% &	4.18\%	& 2.62\% &	1.40\%	& 0.64\% \\
			\hline
			
			Ant Num = $108$ & 4.36\%	& 2.94\% & 1.71\% &	0.87\%	& 0.38\% \\
			\hline
		\end{tabular}
	\end{table}
	Table \ref{tab:ABC} demonstrates that as the ratio between the number of antennas per SAT and the number of served UTs increases, the performance gap diminishes rapidly.
	
\begin{figure}[h]
		\centerline{\includegraphics[scale=0.4]{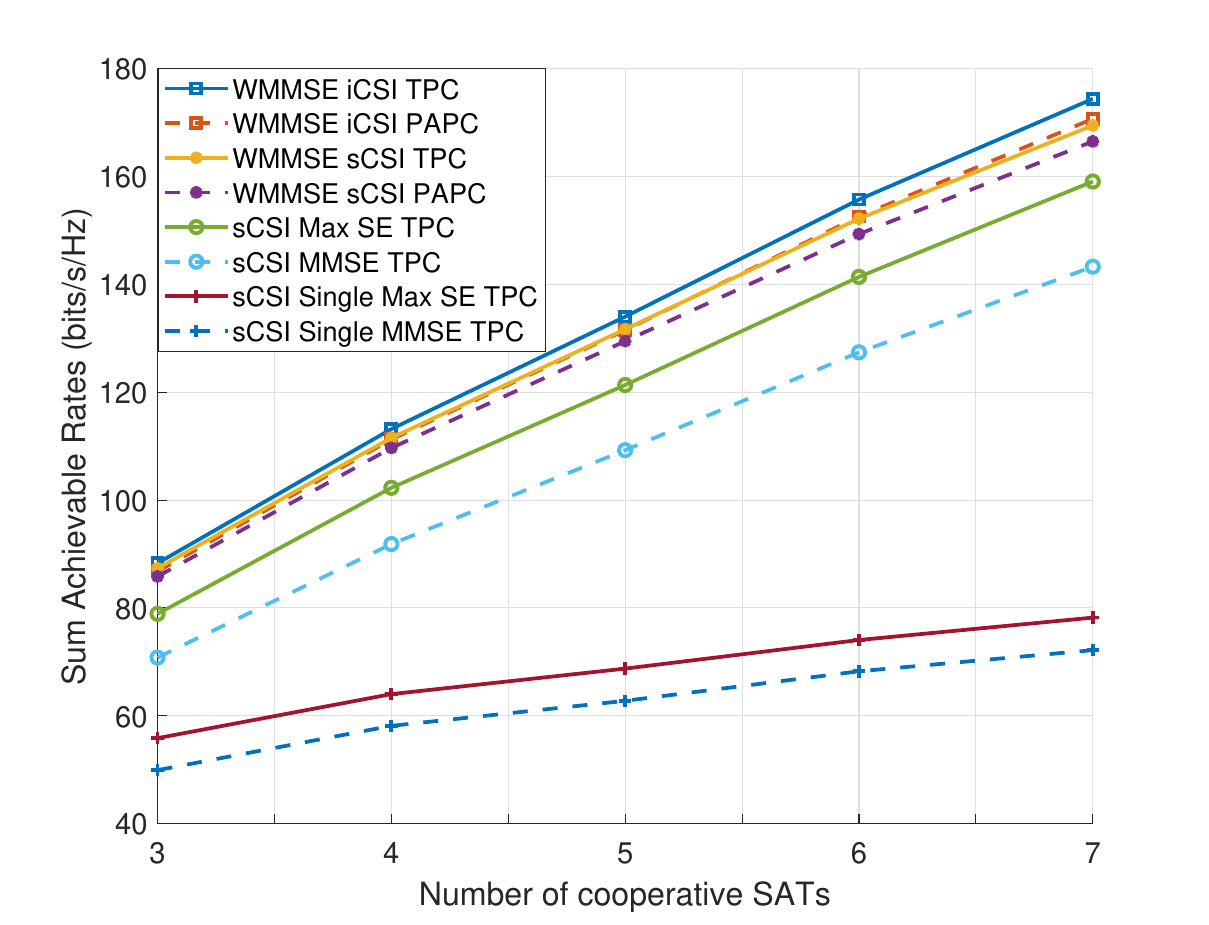}}
		\caption{Performance gap under different numbers of cooperative SATs. $M_x = 6, M_y = 6$; $K = 10$; $P = 15$~dBW; $\kappa_{s,k} = 10$~dB; $\varrho_{s,k}^2 = 0.05$.}
		\label{Fig:Diff_co_SATs_Num}
	\end{figure}

	Fig. \ref{Fig:Diff_co_SATs_Num} illustrates the proposed algorithms' performance varying with the number of cooperative SATs.
	The performance gap remains small between the ``WMMSE iCSI" and ``WMMSE sCSI" algorithms across all SAT configurations.
	Crucially, the performance gap between ``WMMSE sCSI TPC" and ``sCSI Max SE TPC" is highly stable, indicating that the key factor determining the performance gap
	is the number of antennas per SAT, rather than the total number of antennas across all cooperative SATs.
	From Fig. \ref{Fig:Diff_co_SATs_Num}, we also observe that the performance gain of multi-stream transmission over single-stream transmission increases as the number of cooperative SATs increases. This is expected because more cooperative SATs provide a higher number of transmit streams per UT; thus, the performance gain increases accordingly.
	
\begin{figure}[h]
		\centerline{\includegraphics[scale=0.33]{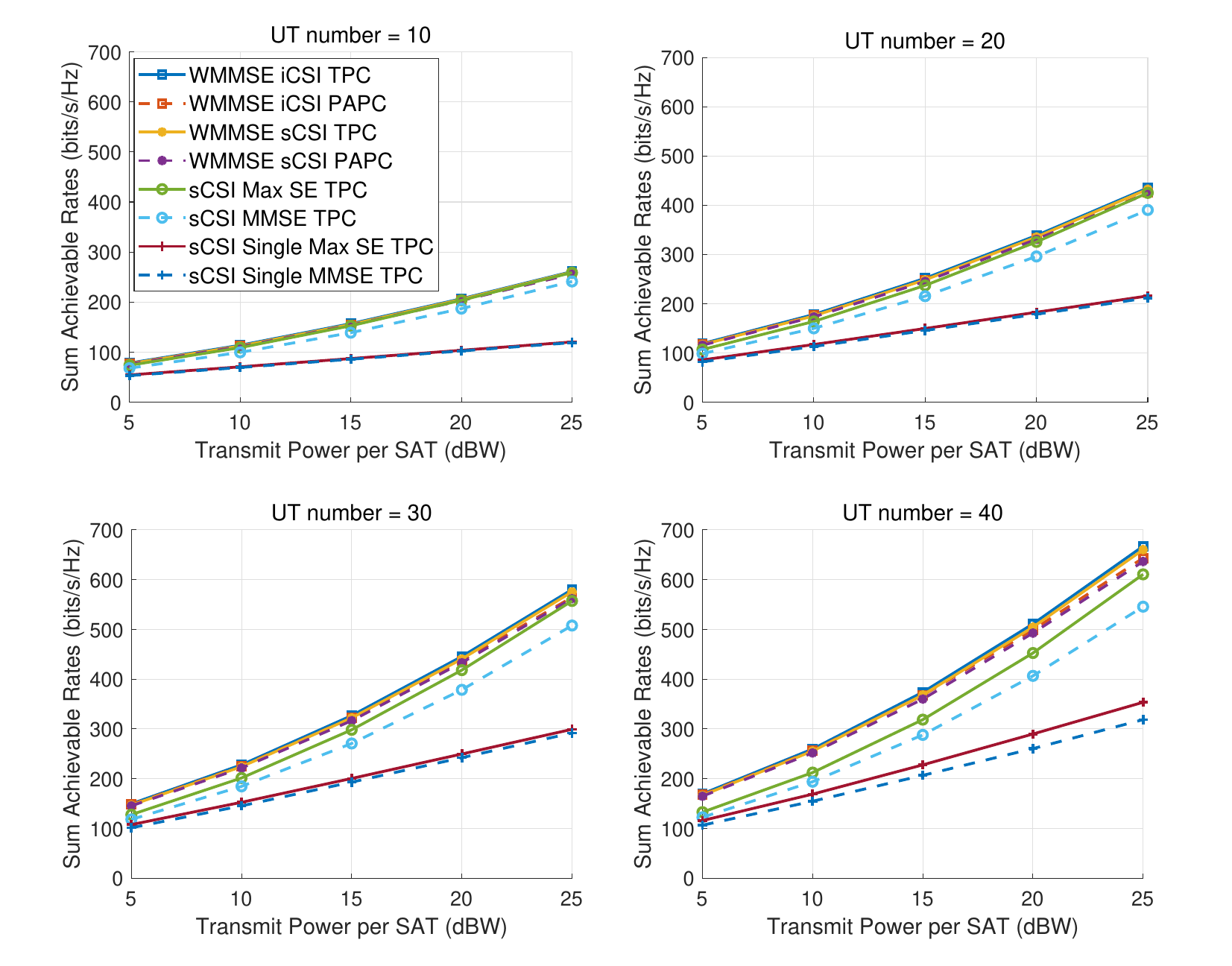}}
		\caption{Performance comparison under different numbers of served UTs. $S = 4$; $M_x = 18, M_y = 6$; $\kappa_{s,k} = 10$~dB; $\varrho_{s,k}^2 = 0.05$.}
		\label{Fig:Diff_UTs_Num}
	\end{figure}

\begin{figure}[h]
		\centerline{\includegraphics[scale=0.38]{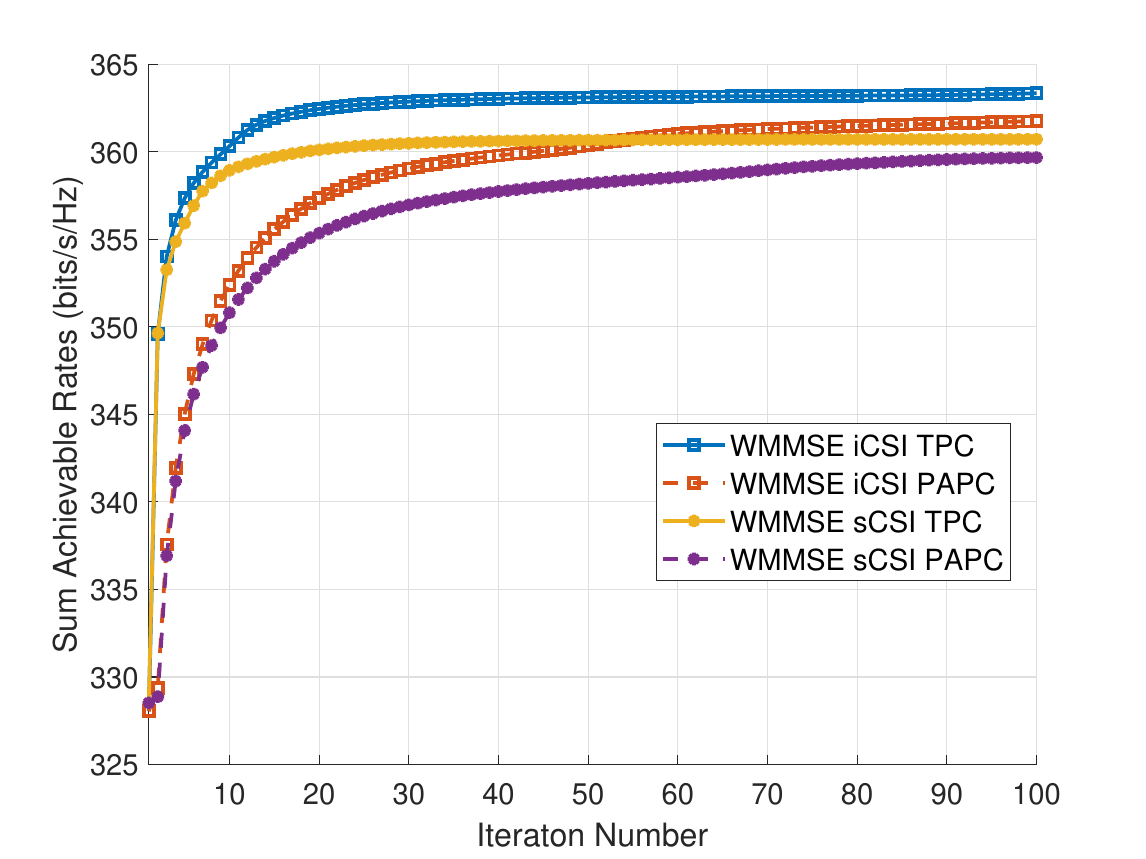}}
		\caption{Convergence behavior of the WMMSE-based algorithms. $S = 4$; $M_x = 18, M_y = 6$; $K = 30$; $P = 15$~dBW; $\kappa_{s,k} = 10$~dB; $\varrho_{s,k}^2 = 0.05$.}
		\label{Fig:Convergence_behavior_optimal_algorithms}
	\end{figure}

	Based on the previous simulation results, we set a relatively large number of antennas per SAT (i.e., $108$) to evaluate the performance of the proposed algorithms against different numbers of served UTs in Fig. \ref{Fig:Diff_UTs_Num}.
	The performance gap between ``WMMSE iCSI" and ``WMMSE sCSI" remains small for all considered numbers of served UTs. On the other hand, the performance gap between ``sCSI Max SE TPC" and the WMMSE sCSI algorithms increases as the number of served UTs increases. Notably, at $K = 30$, the performance gap is still small, corresponding to a ratio of $\frac{M}{K} = 3.6$. As this ratio decreases further, the precoding matrices of ``sCSI Max SE TPC" can serve as a good starting point for the iterative WMMSE algorithms. As detailed later in Fig. \ref{Fig:Convergence_behavior_optimal_algorithms}, executing only the first few iterations can achieve near-converged WMMSE performance.
	
\begin{figure}[h]
		\centerline{\includegraphics[scale=0.35]{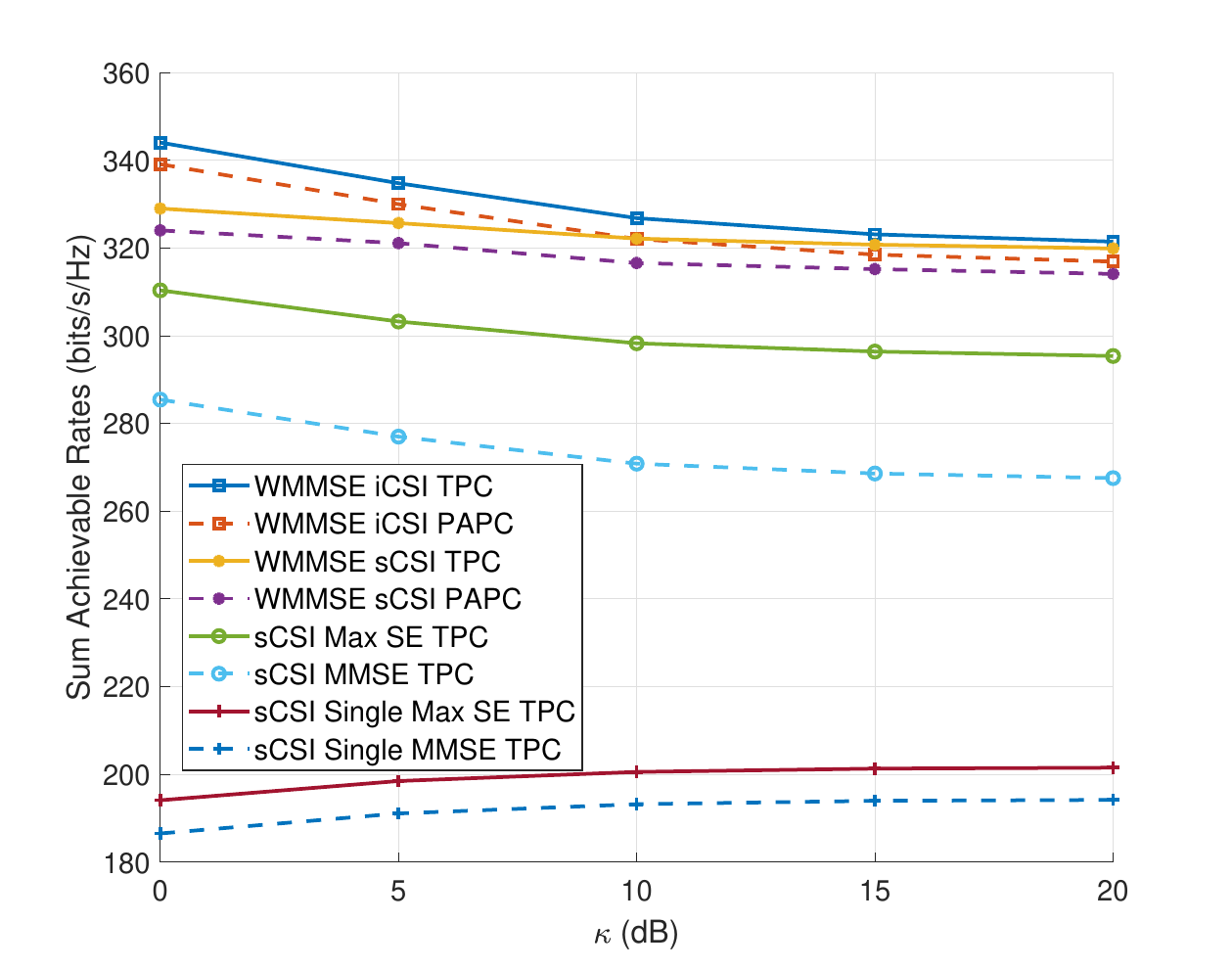}}
		\caption{Effect of Rician factor on the performance of different algorithms. $S = 4$; $M_x = 18, M_y = 6$; $K = 30$; $P = 15$~dBW; $\varrho_{s,k}^2 = 0.05$.}
		\label{Fig:Total_Pow_Per_ANT_Pow_Comparison}
	\end{figure}

	In Fig. \ref{Fig:Total_Pow_Per_ANT_Pow_Comparison}, we investigate the impact of the Rician factor ($\kappa$) on algorithm performance.
	As $\kappa$ decreases, the achievable sum rates increase for the multi-stream algorithms, but decline for the single-stream algorithms. This is because as $\kappa$ decreases, the channel characteristics transition from Rician to Rayleigh fading. The channel energy is spread across multipath components instead of concentrating on the LoS component, which benefits multi-stream transmission but degrades the performance of single-stream transmission. Furthermore, the ``WMMSE sCSI" algorithms perform comparably to the ``WMMSE iCSI" upper bound across the evaluated Rician factor range, though the performance gap widens slightly as $\kappa$ decreases. On the other hand, the performance of ``sCSI Max SE/MMSE TPC" is more robust against the effect of the Rician factor. This is because the inter-UT interference suppression is based on the directional cosines at the SAT, which are independent of the Rician factor.
	
\begin{figure}[h]
		\centerline{\includegraphics[scale=0.36]{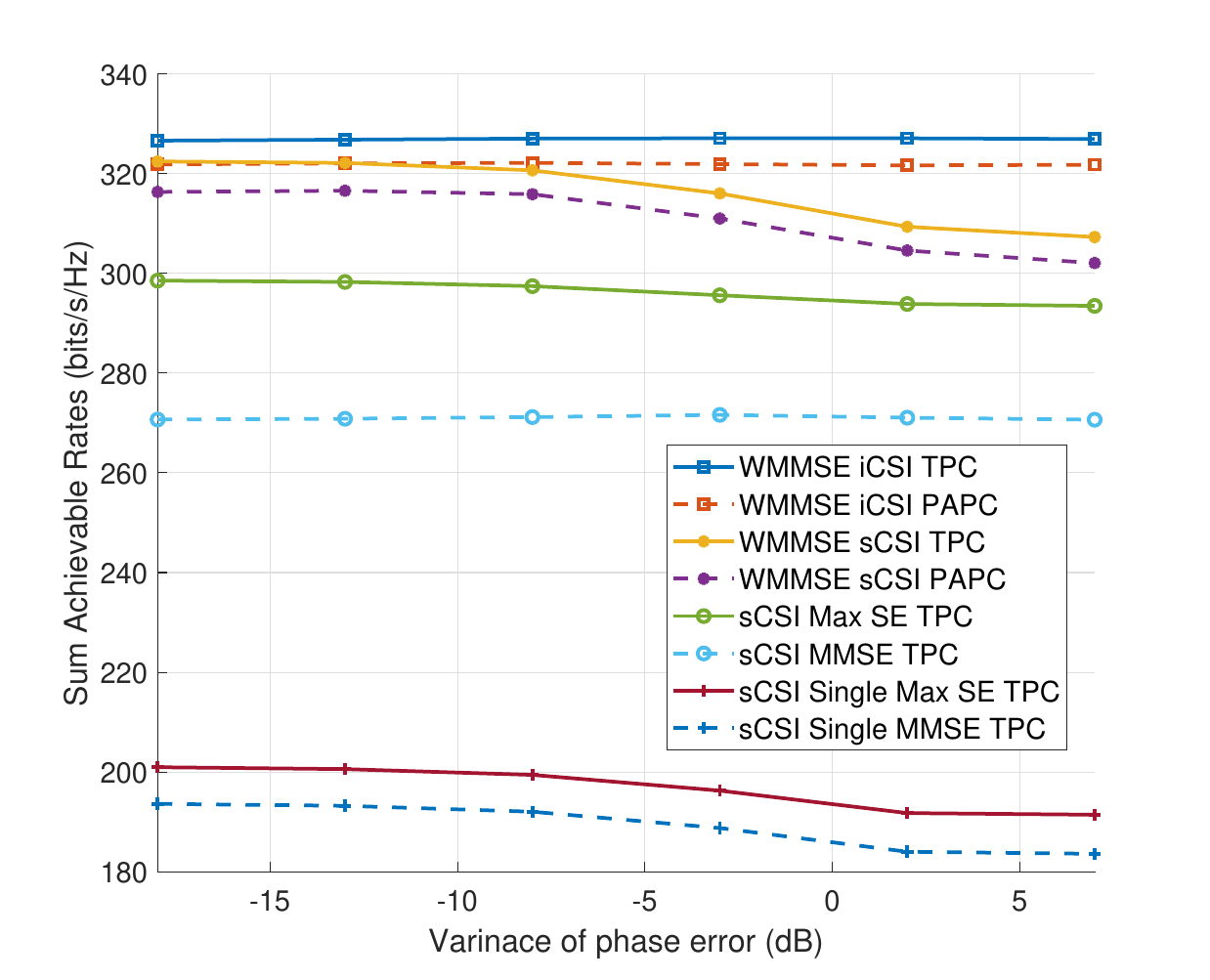}}
		\caption{Effect of variance of phase error on the performance of different algorithms. $S = 4$; $M_x = 18$, $M_y = 6$; $K = 30$; $P = 15$~dBW; $\kappa_{s,k} = 10$~dB.}
		\label{Fig:Effect_phase_error_variance}
	\end{figure}

	Fig. \ref{Fig:Effect_phase_error_variance} also demonstrates the robustness of the ``sCSI Max SE/MMSE TPC" approach against phase error variance.
	As the phase error variance increases, all sCSI-based algorithms experience performance degradation. However, the low-complexity algorithms exhibit a smaller performance degradation. On the other hand, we assume that the phase error for each channel realization is known for the ``WMMSE iCSI" algorithms. Because the precoding matrices can fully compensate for this known phase error, we do not observe any performance degradation in the iCSI case as the phase error variance increases.
	
\begin{figure}[h]
		\centerline{\includegraphics[scale=0.325]{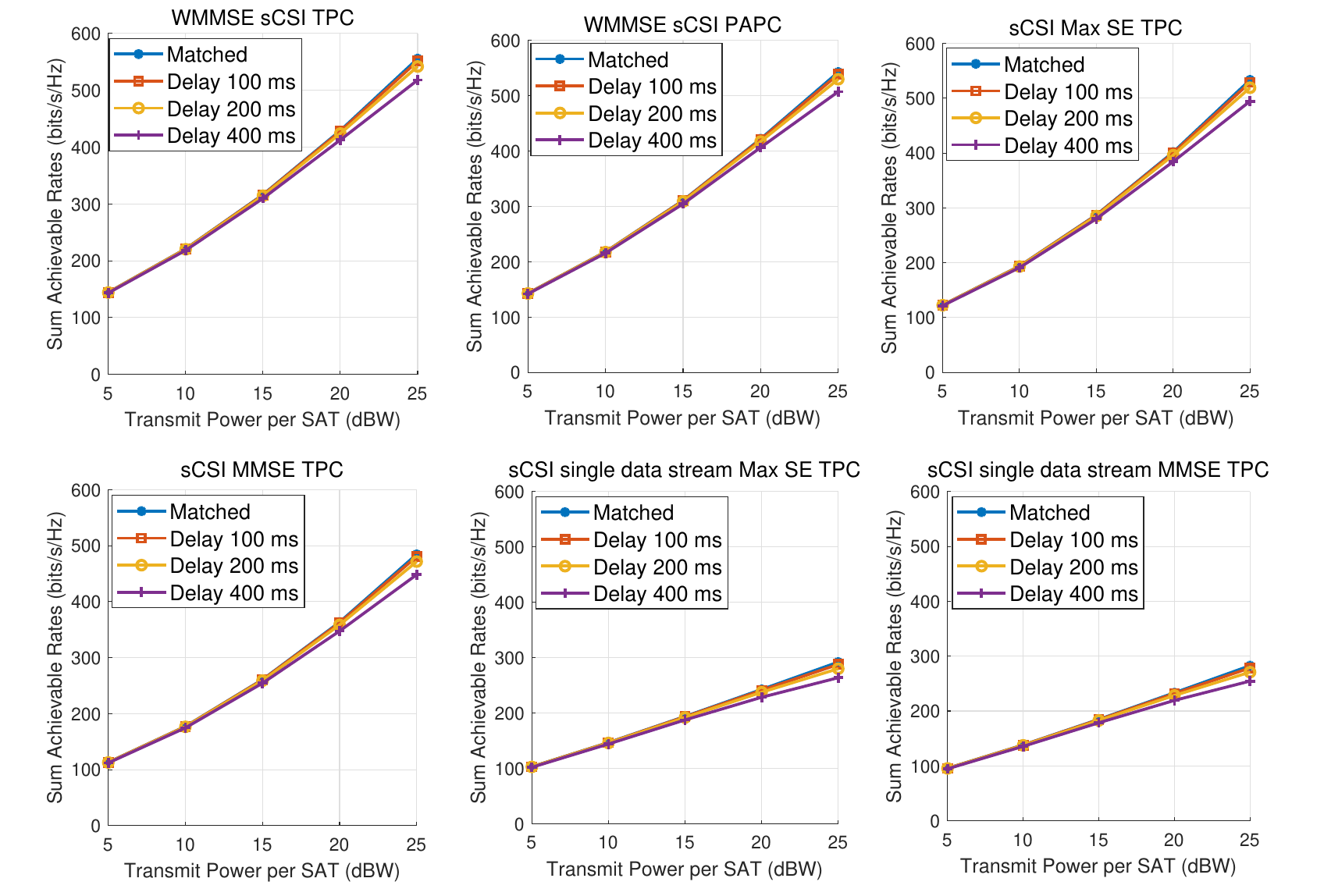}}
		\caption{Effect of delayed sCSI on the performance of the proposed sCSI-based algorithms. $S = 4$; $M_x = 18$, $M_y = 6$; $K = 30$; $\kappa_{s,k} = 10$~dB; $\varrho_{s,k}^2 = 0.05$.}
		\label{Fig:Outdated_CSI_Analysis}
	\end{figure}

	We simulate the effect of delayed sCSI on the performance of sCSI-based algorithms in Fig. \ref{Fig:Outdated_CSI_Analysis}. 
	Delay intervals ranging from 100~ms to 400~ms are considered to evaluate the performance degradation due to mismatched sCSI for precoding and signal propagation. As expected, the sum spectral efficiency gradually decreases as the sCSI staleness increases. However, it is observed that the curves corresponding to the ``Matched" (perfectly synchronized sCSI) and ``Delay 100~ms" cases are almost indistinguishable. This indicates that an update period of 100~ms results in a negligible performance loss.
	
	Fig. \ref{Fig:Convergence_behavior_optimal_algorithms} illustrates the convergence behavior of the iterative WMMSE algorithms.
	The initial precoding matrices for both the TPC and PAPC algorithms are set using the robust precoder generated by Algorithm \ref{algm:sum_rate_max_TPC_sCSI}.
	As shown in Fig. \ref{Fig:Convergence_behavior_optimal_algorithms}, the iterative WMMSE algorithms under TPC converge to a stationary point at around $30$ iterations. On the other hand, the PAPC algorithms require around $100$ iterations. However, since the performance gap for the iterative WMMSE PAPC algorithms between iteration 30 and iteration 100 is only about 1\%, we set the maximum number of iterations to $N_{\text{iter}} = 30$ and $\epsilon = 10^{-3}$ for both the iterative WMMSE TPC and PAPC algorithms in our simulations.
	
\begin{figure}[h]
		\centerline{\includegraphics[scale=0.4]{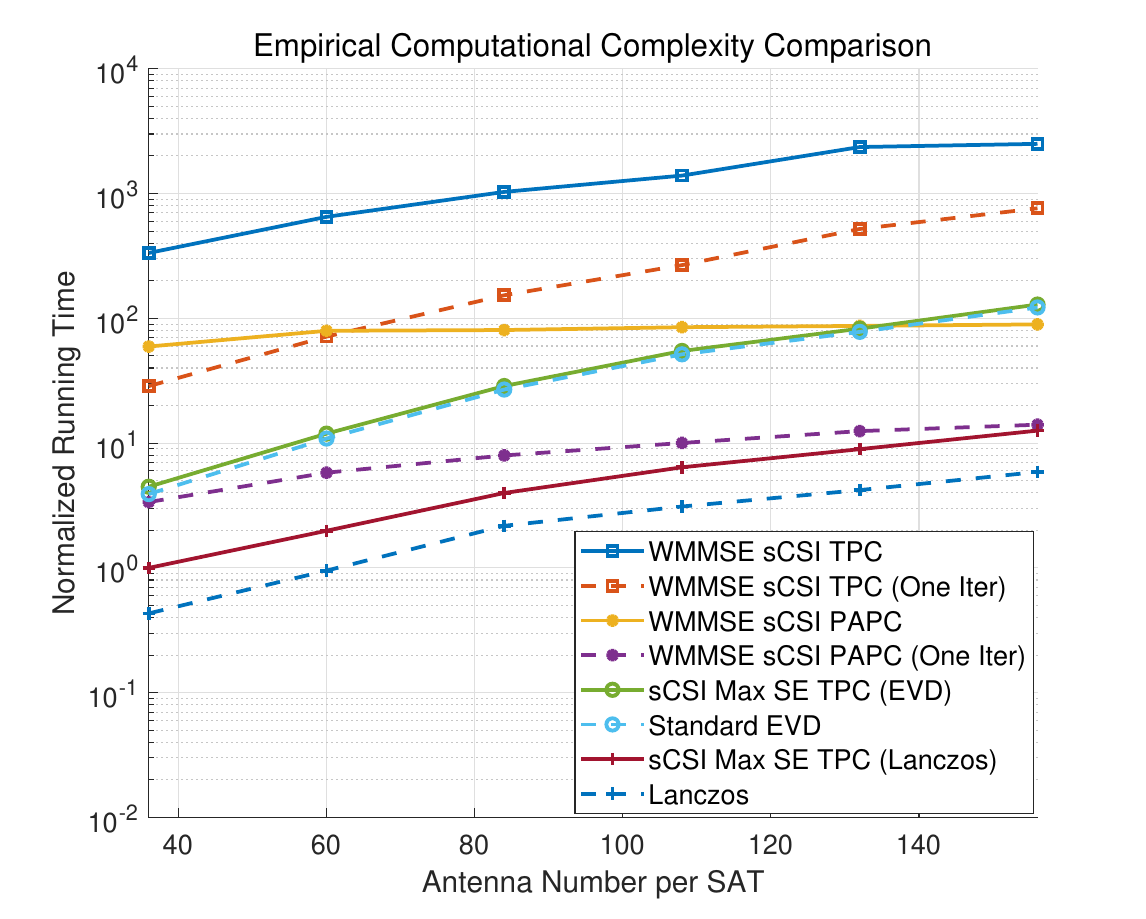}}
		\caption{Running time with respect to the number of antennas per SAT for the proposed sCSI-based algorithms. $S = 4$; $K = 10$; $P = 15$~dBW; $\kappa_{s,k} = 10$~dB; $\varrho_{s,k}^2 = 0.05$.}
		\label{Fig:Running_time_comparison}
	\end{figure}

	While our formal complexity expressions depend on multiple problem dimensions, Fig. \ref{Fig:Running_time_comparison} illustrates how the running times of the proposed sCSI-based algorithms scale with the number of antennas per SAT, providing empirical evidence that aligns with our theoretical analysis.
	Simulations were conducted on an AMD Ryzen 5 7500F 6-Core Processor (3.70 GHz) using MATLAB 2024b. All running times are normalized against that of ``sCSI Max SE TPC (Lanczos)" with $M = 36$.
	Notably, the ``WMMSE sCSI TPC/PAPC" algorithms are initialized using the precoding matrices obtained via ``sCSI Max SE TPC (EVD)". As the ratio between the number of antennas per SAT and the number of served UTs increases, the performance gap between ``sCSI Max SE TPC" and ``WMMSE sCSI TPC/PAPC" decreases. Therefore, we also provide the running times of a single iteration for ``WMMSE sCSI TPC/PAPC", which are labeled as ``One Iter". As depicted in Fig. \ref{Fig:Running_time_comparison}, the computational complexity of ``WMMSE sCSI PAPC" scales linearly with the number of antennas per SAT. Furthermore, we evaluated the running times of ``sCSI Max SE TPC" when the EVD operation is implemented using the standard EVD and the Lanczos algorithm, respectively. The standalone running times of these two operations are also provided. As shown in Fig. \ref{Fig:Running_time_comparison}, a significant complexity reduction is achieved when the standard EVD is replaced with the Lanczos algorithm.
	
\begin{figure}[htbp]
		\centerline{\includegraphics[scale=0.4]{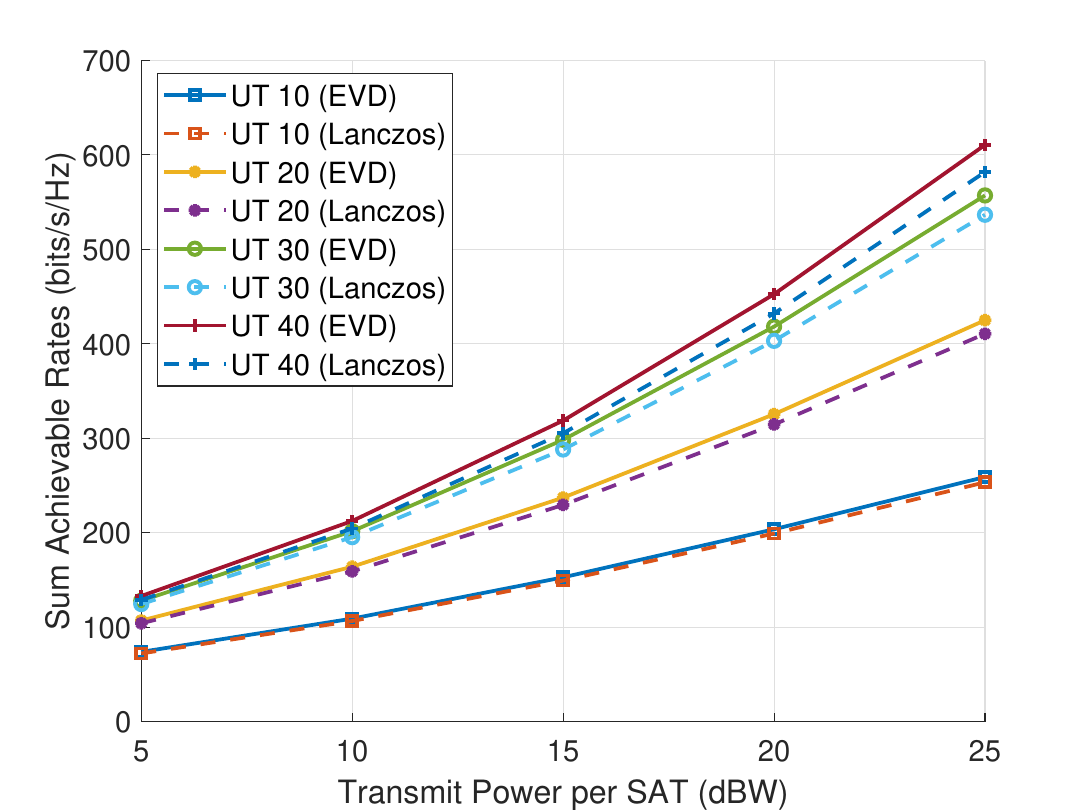}}
		\caption{Performance comparison between applying the Lanczos algorithm and the standard EVD. $S = 4$; $M_x = 18, M_y = 6$; $\kappa_{s,k} = 10$ dB; $\varrho_{s,k}^2 = 0.05$.}
		\label{Fig:Lanczos_simulation}
	\end{figure}

	The performance comparison between ``sCSI Max SE TPC (EVD)" and ``sCSI Max SE TPC (Lanczos)" is provided in Fig. \ref{Fig:Lanczos_simulation}.
	It is observed that the performance gap between applying the Lanczos algorithm and the standard EVD is small under the considered power levels and numbers of UTs, with a maximum gap of less than $5\%$.
	
\begin{figure}[h]
		\centerline{\includegraphics[scale=0.38]{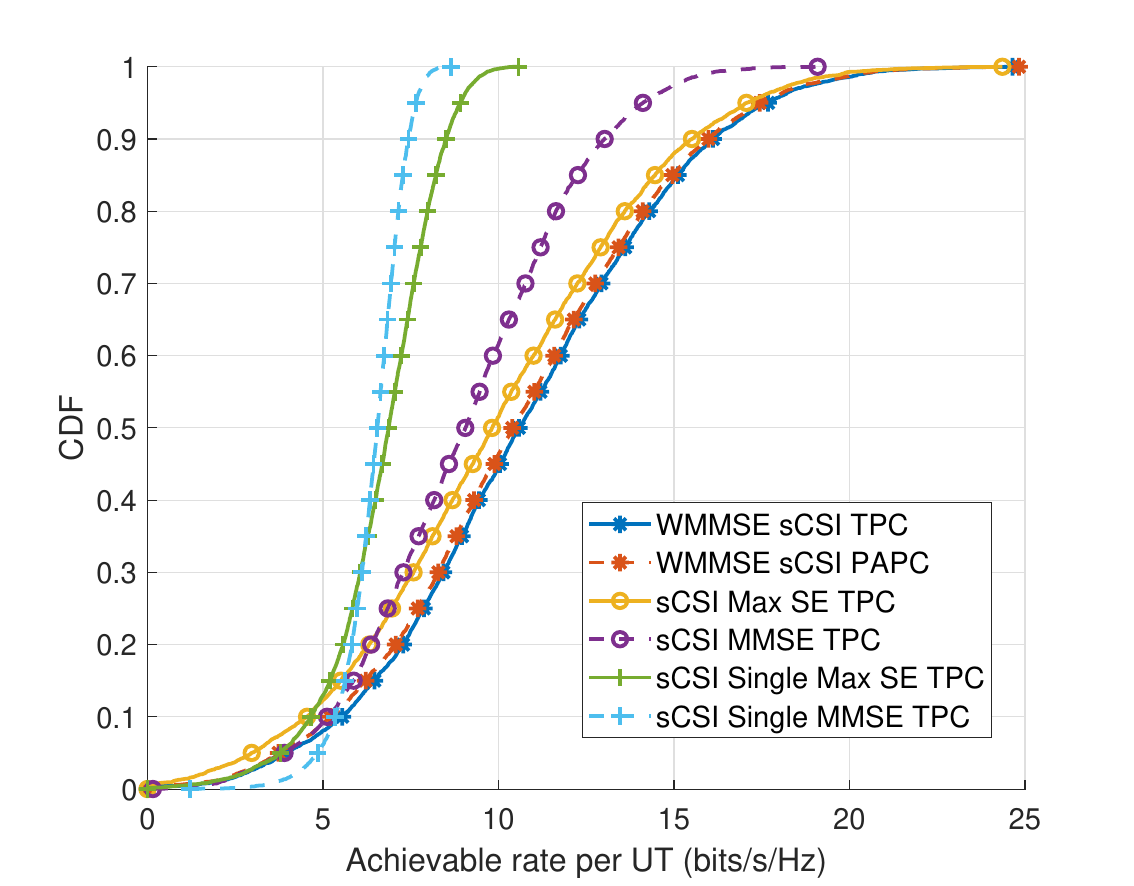}}
		\caption{CDF of proposed sCSI-based algorithms. $S = 4$; $M_x = 18, M_y = 6$; $K = 30$; $P = 15$ dBW; $\kappa_{s,k} = 10$ dB; $\varrho_{s,k}^2 = 0.05$.}
		\label{Fig:CDF_Analysis}
	\end{figure}

	Fig. \ref{Fig:CDF_Analysis} plots the cumulative distribution function (CDF) of the achievable rate per UT for the sCSI-based algorithms. 
	In particular, the $5\%$ SE values per UT for ``iterative WMMSE sCSI TPC/PAPC" are 3.89 bps/Hz and 3.73 bps/Hz, respectively. For ``sCSI Max SE/MMSE TPC", they are 2.96 bps/Hz and 3.90 bps/Hz, respectively. In the case of single-stream transmission, the values are 3.78 bps/Hz and 4.85 bps/Hz for ``sCSI Single Max SE/MMSE TPC", respectively. These results indicate that single-stream transmission outperforms multi-stream transmission, and the MMSE criterion also outperforms the Max SE criterion in terms of the $5\%$ SE per UT. This reveals an inherent trade-off: while the proposed multi-stream approach is effective for maximizing average performance and robustness, it is not uniformly superior across all user-centric fairness criteria.
	To further improve the fairness among different UTs, we can consider assigning larger weights to UTs experiencing poor channel conditions \cite{Shi2011WMMSE}.
	
	\section{Conclusion}
	\label{sec:conclusion}
	We have demonstrated the feasibility and performance benefits of SAT cooperation in DL multi-stream, multi-SAT mMIMO systems. Iterative WMMSE precoding designs using iCSI and sCSI under TPC and PAPC conditions are provided. Furthermore, a robust and low-complexity precoder design under the TPC condition is also developed. By utilizing the Lanczos algorithm, the computational complexity of this design is significantly reduced, making it highly suitable for practical applications. Notably, when the ratio between the number of antennas per SAT and the number of served UTs is large—as is typical in the mMIMO regime—the proposed robust and low-complexity method achieves performance close to the local optimum. Finally, the precoding matrices generated by the proposed low-complexity method can also serve as an effective starting point for the iterative WMMSE algorithms to facilitate convergence. Importantly, while the proposed designs excel in average performance and robustness, they are not uniformly superior across all user-centric fairness metrics. Weighted approaches \cite{Shi2011WMMSE} can be incorporated when considering different fairness requirements.

	\appendices
	\section{}
	\label{sec:sCSI_info}
	We first define $\mathbf{\Pi} = \mathbf{W}_k\mathbf{W}_k^H + \sum_{k'\neq k}^K\tilde{\mathbf{W}}_{k'}\tilde{\mathbf{W}}_{k'}^H$ and $\mathbf{\Pi}_k = \sum_{k'\neq k}^K\tilde{\mathbf{W}}_{k'}\tilde{\mathbf{W}}_{k'}^H$. Then $r_k$ in (\ref{eq:r_k}) can be written as:
	\begin{align}
		r_k& = \log\det\Big(\big(\hat{\mathbf{H}}_k\mathbf{\Pi}\hat{\mathbf{H}}_k^H + \sigma_k^2 \mathbf{I}_N\big)\big(\hat{\mathbf{H}}_k\mathbf{\Pi}_k\hat{\mathbf{H}}_k^H+ \sigma_k^2 \mathbf{I}_N\big)^{-1}\Big) \nonumber \\
		&= \log\det\Big(\hat{\mathbf{H}}_k\mathbf{\Pi}\hat{\mathbf{H}}_k^H + \sigma_k^2 \mathbf{I}_N\Big)\nonumber \\
		&\quad-\log\det\Big(\hat{\mathbf{H}}_k\mathbf{\Pi}_k\hat{\mathbf{H}}_k^H+ \sigma_k^2 \mathbf{I}_N\Big)\nonumber \\
		&= \log\det\Big(\hat{\mathbf{H}}_k^H \hat{\mathbf{H}}_k\mathbf{\Pi}+ \sigma_k^2 \mathbf{I}_{SM}\Big) \nonumber \\
		&\quad-\log\det\Big(\hat{\mathbf{H}}_k^H\hat{\mathbf{H}}_k\mathbf{\Pi}_k + \sigma_k^2 \mathbf{I}_{SM}\Big).
		\label{eq:r_k_2}
	\end{align}
	Therefore, $\mathbb{E}\{r_k\}$ can be expressed as:
	\begin{align}
		\mathbb{E}\{r_k\} = & \mathbb{E}\{\log\det\big(\hat{\mathbf{H}}_k^H \hat{\mathbf{H}}_k\mathbf{\Pi}+ \sigma_k^2 \mathbf{I}_{SM}\big)\}  \nonumber \\ &-\mathbb{E}\{\log\det\big(\hat{\mathbf{H}}_k^H\hat{\mathbf{H}}_k\mathbf{\Pi}_k + \sigma_k^2 \mathbf{I}_{SM}\big)\}.
		\label{eq:E_r_k}
	\end{align}
	By applying Jensen's inequality to (\ref{eq:E_r_k}), we have
	\begin{align}
		\mathbb{E}\{r_k\} \approx & \log\det(\mathbb{E}\{\hat{\mathbf{H}}_k^H\hat{\mathbf{H}}_k\}\mathbf{\Pi} + \sigma_k^2\mathbf{I}_{SM}) \nonumber \\
		&-\log\det\big(\mathbb{E}\{\hat{\mathbf{H}}_k^H\hat{\mathbf{H}}_k\}\mathbf{\Pi}_k + \sigma_k^2 \mathbf{I}_{SM}\big) \nonumber \\
		= & \log\det(\tilde{\mathbf{H}}_k^H\tilde{\mathbf{H}}_k\mathbf{\Pi} + \sigma_k^2\mathbf{I}_{SM}) \nonumber \\
		&-\log\det\big(\tilde{\mathbf{H}}_k^H\tilde{\mathbf{H}}_k\mathbf{\Pi}_k + \sigma_k^2 \mathbf{I}_{SM}\big).
		\label{eq:E_r_k_2}
	\end{align}
	Note that the expression for $\mathbb{E}\{r_k\}$ in (\ref{eq:E_r_k_2}) is equivalent to (\ref{eq:E_r_k_3}). On the other hand, when we take the gradient of $\mathbb{E}\{\text{MSE}_k\}$ with respect to $\tilde{\mathbf{F}}_k$ and set it to $\mathbf{0}$, we obtain the optimal form of $\tilde{\mathbf{F}}_k$ as given in (\ref{eq:tilde_F_opt}).
	By substituting $\tilde{\mathbf{F}}_k^{\text{opt}}$ into $\mathbb{E}\{\text{MSE}_k\}$, we obtain $\mathbb{E}\{\text{MMSE}_k\}$:
	\begin{equation}
		\mathbb{E}\{\text{MMSE}_k\} \approx (\mathbf{I}_{d_k} + \mathbf{W}_k^H\tilde{\mathbf{H}}_k^H(\tilde{\mathbf{R}}'_z)^{-1}\tilde{\mathbf{H}}_k\mathbf{W}_k)^{-1}.
		\label{eq:E_MMSE_k}
	\end{equation}
	It can be readily shown that
	\begin{equation}
		\mathbb{E}\{r_k\} = \log\det(\mathbb{E}\{\text{MMSE}_k\}^{-1}).
	\end{equation}
	So far, we have constructed the closed-form expressions for $\mathbb{E}\{r_k\}$, $\mathbb{E}\{\text{MMSE}_k\}$, $\mathbb{E}\{\text{MSE}_k\}$, and $\tilde{\mathbf{F}}_k^{\text{opt}}$. Therefore, the locally optimal precoding matrix $\mathbf{W}_k$ under the sCSI condition can be calculated using the WMMSE framework. 
	\hfill $\blacksquare$
	
	\section{}
	We first review some aspects of majorization theory that will be used to prove Theorem \ref{theorem:eigenvalues} later.
	For any vector $\mathbf{x} \in \mathbb{R}^n$, let $[\mathbf{x}]_{1} \geq \cdots \geq [\mathbf{x}]_{n}$
	denote the components of $\mathbf{x}$ in decreasing order (also known as the order statistics of $\mathbf{x}$).
	\begin{definition}
		\cite{Marshall1979Inequalities} Let $\mathbf{x}, \mathbf{y} \in \mathbb{R}^n$. A vector $\mathbf{x}$ is \textit{majorized} by a vector $\mathbf{y}$ (or $\mathbf{y}$ \textit{majorizes} $\mathbf{x}$) if
		\begin{gather}
			\sum_{i=1}^k [\mathbf{x}]_{i} \leq \sum_{i=1}^k [\mathbf{y}]_{i},\; 1\leq k \leq n-1;\quad
			\sum_{i=1}^n [\mathbf{x}]_{i} = \sum_{i=1}^n [\mathbf{y}]_{i}. \nonumber
		\end{gather}
		This relationship is denoted by $\mathbf{x} \prec \mathbf{y}$.
		\label{definition:majorization}
	\end{definition}
	\begin{definition}
		\cite{Marshall1979Inequalities}
		A real-valued function $\phi$ defined on a set $\mathcal{A} \subseteq \mathbb{R}^{n}$ is said to be Schur-concave on $\mathcal{A}$ if $\mathbf{x} \prec \mathbf{y}$ on $\mathcal{A} \Rightarrow \phi(\mathbf{x}) \geq \phi(\mathbf{y})$. 
		\label{Def:Schur-concave}
	\end{definition}
	\begin{lemma}
		\cite{Marshall1979Inequalities} Let $\mathbf{R}$ be an $n \times n$ Hermitian matrix with diagonal elements 
		$\mathbf{d}$ and eigenvalues 
		$\bm{\lambda}$. Then $\mathbf{d} \prec \bm{\lambda}.$
		\label{lemma:d_lambda}
	\end{lemma}
	
	\section{}
	\label{appdix:Theorem1proof}
	It is sufficient to show that there exists a diagonal matrix $\mathbf{P}$ such that for $\tilde{\mathbf{A}} = \mathbf{U}_2\mathbf{P}^{1/2}$ satisfies the following two conditions: $\mathbf{\Lambda}_1 = \tilde{\mathbf{\Lambda}}_2\text{diag}(\tilde{\mathbf{p}})$ and $\text{tr}(\tilde{\mathbf{A}}\tilde{\mathbf{A}}^H) \leq \text{tr}(\mathbf{A}\mathbf{A}^H)$, where $\tilde{\mathbf{\Lambda}}_2$ is the upper-left $s \times s$ block of $\mathbf{\Lambda}_2$ (i.e., $\tilde{\mathbf{\Lambda}}_2 = [\mathbf{\Lambda}_2]_{1:s,1:s}$).
	First, we note that
	\begin{align}
		& \text{tr}(\mathbf{A}^H\mathbf{\Psi}\mathbf{A}) = \, \text{tr}(\mathbf{A}^H\mathbf{U}_2\mathbf{\Lambda}_2\mathbf{U}_2^H\mathbf{A}) = \text{tr}(\bar{\mathbf{A}}\mathbf{\Lambda}_2\bar{\mathbf{A}}^H) \nonumber \\ 
		& = \, \text{tr} \Big(\sum_{i=1}^n\lambda_2^i\bar{\mathbf{a}}_i\bar{\mathbf{a}}_i^H\Big)
		= \sum_{i=1}^n\text{tr}(\lambda_2^i\bar{\mathbf{a}}_i\bar{\mathbf{a}}_i^H) 
		= \sum_{i=1}^n \lambda_2^i\|\bar{\mathbf{a}}_i\|^2,
		\label{eq:APsiA} 
	\end{align}
	where $\bar{\mathbf{A}}=\mathbf{A}^H\mathbf{U}_2$ and $\bar{\mathbf{a}}_i$ is the $i$-th column of $\bar{\mathbf{A}}$.
	Since in the formulation of the Theorem we assumed $\mathbf{\Psi} = \mathbf{U}_2 \mathbf{\Lambda}_2 \mathbf{U}_2^H$ and $\tilde{\mathbf{A}} = \mathbf{U}_2\mathbf{P}^{1/2}$, we obtain
	\begin{equation}
		\text{tr}(\tilde{\mathbf{A}}^H\mathbf{\Psi}\tilde{\mathbf{A}}) = \text{tr}((\mathbf{P}^{1/2})^T\mathbf{\Lambda}_2\mathbf{P}^{1/2}) = \sum_{i=1}^{s}\lambda_2^i p_i,
	\end{equation}
	where $p_i = [\tilde{\mathbf{p}}]_i$. We also have
	\begin{align}
		\tilde{\mathbf{A}}^H\mathbf{\Psi} \tilde{\mathbf{A}}  = (\mathbf{U}_2 \mathbf{P}^{1/2})^H(\mathbf{U}_2 \mathbf{\Lambda}_2 \mathbf{U}_{2}^H) (\mathbf{U}_2 \mathbf{P}^{1/2}) 
		= \tilde{\mathbf{\Lambda}}_2\text{diag}(\tilde{\mathbf{p}}).
	\end{align}
	We require that $\tilde{\mathbf{\Lambda}}_2\text{diag}(\tilde{\mathbf{p}}) = \mathbf{\Lambda}_1$. Since $\text{diag}(\tilde{\mathbf{p}})$, $\mathbf{\Lambda}_1$, and $\mathbf{\Lambda}_2$ are diagonal, $p_i$ must satisfy $\lambda_1^i = \lambda_2^ip_i, i=1,\dots,s$. Recall that for a Hermitian $\mathbf{R}$, the sum of its eigenvalues is equal to $\text{tr}(\mathbf{R})$. Thus, we have
	\begin{equation}
		\sum_{i=1}^n\lambda_2^i\|\bar{\mathbf{a}}_i\|^2 = \sum_{i=1}^{s}\lambda_1^i = \sum_{i=1}^{s}\lambda_2^ip_i. 
	\end{equation}
	To be consistent with the following derivation, we also have 
	\begin{equation}
		\sum_{i=1}^n\lambda_2^i\|\bar{\mathbf{a}}_i\|^2 = \sum_{i=1}^{n}\lambda_1^i = \sum_{i=1}^{n}\lambda_2^ip_i,
	\end{equation}
	where $\lambda_1^i=0$ and $p_i = 0, i = s+1,\dots,n$.
	
	We define $\mathbf{x} = [\lambda_2^1\|\bar{\mathbf{a}}_1\|^2,\dots,\lambda_2^n\|\bar{\mathbf{a}}_n\|^2]$. Note that the elements of $\mathbf{x}$ are not necessarily in decreasing order. For example, we can have $\lambda_2^i\|\bar{\mathbf{a}}_i\|^2 > \lambda_2^1\|\bar{\mathbf{a}}_1\|^2, i \neq 1$. In this case, we can choose $q_1$ and $q_i$ satisfying the following condition:
	\begin{equation}
		\lambda_2^i\|\bar{\mathbf{a}}_i\|^2 = \lambda_2^1 q_1 = [\mathbf{x}]_{i} \text{ and } \lambda_2^1\|\bar{\mathbf{a}}_1\|^2 = \lambda_2^i q_i = [\mathbf{x}]_{1}.
	\end{equation}
	Since $[\mathbf{x}]_{i} > [\mathbf{x}]_{1}$, we have
	\begin{equation}
		\frac{[\mathbf{x}]_{i}}{\lambda_2^i}-\frac{[\mathbf{x}]_{i}}{\lambda_2^1} > \frac{[\mathbf{x}]_{1}}{\lambda_2^i} - \frac{[\mathbf{x}]_{1}}{\lambda_2^1},
	\end{equation}
	which indicates that $\|\bar{\mathbf{a}}_1\|^2 + \|\bar{\mathbf{a}}_i\|^2 > q_1 + q_i$. Thus, we can construct a vector $\mathbf{z} = [\lambda_2^1 q_1,\dots,\lambda_2^n q_n]$ whose elements are the same as those of $\mathbf{x}$ but arranged in decreasing order, and which also satisfies $\sum_{i=1}^n q_i \leq \sum_{i=1}^n \|\bar{\mathbf{a}}_i\|^2$. 
	
	We define $\mathbf{y} \triangleq [\lambda_2^1 p_1,\dots, \lambda_2^n p_n]$ and $\mathbf{Q}\triangleq\bar{\mathbf{A}}_2^H\bar{\mathbf{A}}_2$, where $\bar{\mathbf{A}}_2 = \bar{\mathbf{A}}\mathbf{\Lambda}_2^{1/2}$. From the definition of $\mathbf{Q}$, $\mathbf{x}$ is composed of diagonal elements of $\mathbf{Q}$, and $\mathbf{z}$ is composed of diagonal elements of $\mathbf{Q}$ in decreasing order. We further define $\mathbf{T} \triangleq \bar{\mathbf{A}}_2\bar{\mathbf{A}}_2^H$. From the definition of $\mathbf{Q}$ and $\mathbf{T}$, the nonzero eigenvalues of $\mathbf{Q}$ and $\mathbf{T}$ are the same since in both cases they are the squares of the singular values of $\bar{\mathbf{A}}_2$. According to (\ref{eq:APsiA}), we have 
	\begin{equation}
		\mathbf{T} = \bar{\mathbf{A}}\mathbf{\Lambda}_2\bar{\mathbf{A}}^H = \mathbf{A}^H\mathbf{\Psi}\mathbf{A} = \mathbf{U}_1\mathbf{\Lambda}_1\mathbf{U}_1^H.
	\end{equation}
	Thus, the eigenvalues of $\mathbf{T}$ are $\lambda_1^i, i = 1, \dots, s$, and the eigenvalues of $\mathbf{Q}$ are $\lambda_1^i, i = 1, \dots, n$, respectively. Since we set $p_i$ such that $\lambda_2^ip_i = \lambda_1^i, i = 1, \dots, s$ and $p_i = 0, i = s+1, \dots, n$, the vector $\mathbf{y} = [\lambda_2^1 p_1, \dots, \lambda_2^n p_n]$ becomes $[\lambda_1^1, \dots, \lambda_1^n]$. This vector is composed of the eigenvalues of $\mathbf{Q}$.
	
	From Lemma \ref{lemma:d_lambda}, $\mathbf{z} \prec \mathbf{y}$, which indicates that $[\mathbf{y}]_{1} \geq [\mathbf{z}]_{1}$. To satisfy the condition $\sum_{i=1}^n [\mathbf{y}]_{i} = \sum_{i=1}^n [\mathbf{z}]_{i}$, this requires an index subset $\{i\}$ with $[\mathbf{y}]_{i} \leq [\mathbf{z}]_{i}$ to compensate for the gap $[\mathbf{y}]_{1} - [\mathbf{z}]_{1}$ (the compensation may not be precise). We then probably have another index $j$ with $[\mathbf{y}]_{j} \geq [\mathbf{z}]_{j}$, this necessitates another index subset $\{i\}$ with $[\mathbf{y}]_{i} \leq [\mathbf{z}]_{i}$ to compensate for the gap $[\mathbf{y}]_{j} - [\mathbf{z}]_{j}$. In general, due to the constraint $\sum_{i=1}^k [\mathbf{z}]_{i} \leq \sum_{i=1}^k [\mathbf{y}]_{i}, 1\leq k \leq n-1$, the index $i$ with $[\mathbf{y}]_{i} \leq [\mathbf{z}]_{i}$ to compensate for the gap $[\mathbf{y}]_{j} - [\mathbf{z}]_{j}$ is larger than $j$. On the other hand, we can certainly have an index subset $\{i\}$ with $[\mathbf{y}]_{i} \geq [\mathbf{z}]_{i}$, and the gap of $\sum_{i \in \{i\}}([\mathbf{y}]_{i} - [\mathbf{z}]_{i})$ can be compensated by a single index $j$ with $[\mathbf{y}]_{j} \leq [\mathbf{z}]_{j}$. Again, due to the constraints in Definition \ref{definition:majorization}, the index $j$ must be larger than the indices in $\{i\}$. 
	
	We assume $\lambda_2^ip_i \geq \lambda_2^i q_i$ for $i\in \mathcal{S}_1$ and $\lambda_2^ip_i < \lambda_2^iq_i$ for $i \in \mathcal{S}_2$.
	Since $\sum_{i=1}^n\lambda_2^iq_i = \sum_{i=1}^n\lambda_2^ip_i$,
	\begin{equation}
		\sum_{i\in\mathcal{S}_1}\lambda_2^i(p_i - q_i) = \sum_{i\in\mathcal{S}_2}\lambda_2^i(q_i - p_i).
	\end{equation} 
	According to the explanation in the previous paragraph, the indices in $\mathcal{S}_1$ are, on average, smaller than the indices in $\mathcal{S}_2$.
	Since $\{\lambda_2^i\}$ are in decreasing order with respect to $i$, the values $\{\lambda_2^i\}$ for $i\in\mathcal{S}_1$ are, on average, larger than the values $\{\lambda_2^i\}$ for $i\in\mathcal{S}_2$. 
	Thus, we obtain 
	$\sum_{i=1}^n p_i \leq \sum_{i=1}^n q_i \leq \sum_{i=1}^n\|\bar{\mathbf{a}}_i\|^2.$
	Note that
	\begin{align}
		\sum_{i=1}^n\|\bar{\mathbf{a}}_i\|^2 =  \text{tr}(\mathbf{A}^H\mathbf{A}) ,\;
		\sum_{i=1}^n p_i =  \text{tr}(\tilde{\mathbf{A}}^H\tilde{\mathbf{A}}).
	\end{align}
	Thus, we have $\text{tr}(\tilde{\mathbf{A}}\tilde{\mathbf{A}}^H) \leq \text{tr}(\mathbf{A}\mathbf{A}^H)$. The proof is complete.
	
	\section{}
	\label{appdix:Max_SE_problem_conversion}
	Due to the concavity of $\log_2(\cdot)$, we apply Jensen's inequality to the objective function in (\ref{eq:sum_rate_maximization_stats}) to establish the upper bound of $\mathbb{E}\{\log_2(1 + \text{SINR}_{k,i}^{\text{MMSE}})\}$ as $\log_2(\mathbb{E}\{1 + \text{SINR}_{k,i}^{\text{MMSE}}\})$.
	Because $\log_2(\cdot)$ is a monotonically increasing function, maximizing this upper bound allows us to formulate a surrogate objective for (\ref{eq:sum_rate_maximization_stats}) as:
	\begin{gather}
		\underset{\{\mathbf{W}_k\}}{\text{max}}\prod_{k=1}^{K}\prod_{i=1}^{d_k}\mathbb{E}\{1 + \text{SINR}_{k,i}^{\text{MMSE}}\} \;\;
		\text{s.t. } \sum_{k=1}^K \text{tr}\big(\mathbf{W}_k\mathbf{W}_k^H\big) \leq P.
		\label{eq:sum_rate_maximization_stats_2}
	\end{gather}
	Furthermore, from the relationship between SINR and MMSE, we have:
	\begin{equation}
		\mathbb{E}\{1 + \text{SINR}_{k,i}^{\text{MMSE}}\} = \mathbb{E}\{[\text{MMSE}_k]_{i,i}^{-1}\}.
		\label{eq:SINR_MMSE_Equivalence}
	\end{equation} 
	By the definition of MSE, $[\text{MMSE}_k]_{i,i}> 0, \forall i, k$; therefore, $[\text{MMSE}_k]_{i,i}^{-1}$ is a convex function of $[\text{MMSE}_k]_{i,i}$. Applying Jensen's inequality again to (\ref{eq:SINR_MMSE_Equivalence}), we obtain the following relationship: $\mathbb{E}\big\{[\text{MMSE}_k]_{i,i}^{-1}\big\} \geq (\mathbb{E}\{[\text{MMSE}_k]_{i,i}\})^{-1}$.
	Substituting this lower bound into the objective, problem (\ref{eq:sum_rate_maximization_stats_2}) is further approximated as
	\begin{gather}
		\underset{\{\mathbf{W}_k\}}{\text{max}}\prod_{k=1}^{K}\prod_{i=1}^{d_k}\frac{1}{\mathbb{E}\{[\text{MMSE}_k]_{i,i}\}} \;\;\;
		\text{s.t. } \sum_{k=1}^K \text{tr}\big(\mathbf{W}_k\mathbf{W}_k^H\big) \leq P.
		\label{eq:max_sum_rate_equi}
	\end{gather}
	Finally, we apply the approximation in (\ref{eq:MSE_lower_bound}) by replacing $\mathbb{E}\{[\text{MMSE}_k]_{i,i}\}$ with $[\mathbf{E}_k]_{i,i}$ in (\ref{eq:max_sum_rate_equi}). We also incorporate the precoding structure $\mathbf{W}_k = \mathbf{V}_k\mathbf{A}_k$ to suppress the inter-UT interference. With these substitutions, problem (\ref{eq:max_sum_rate_equi}) yields the approximate formulation in (\ref{eq:sum_rate_maximization_stats_3}), which completes the proof.

	\section{}
	\label{Appdix:Schur-Concave}
	We define $\phi(\mathbf{x}) \triangleq \log(-f_0(\mathbf{x})) = \sum_{i}g(x_i)$, where $g(x) = \log(x^{-1})$. 
	The function $\phi$ is Schur-convex because $g(x)$ is a strictly convex function \cite[3.C.1]{Marshall1979Inequalities}. 
	Furthermore, since $f_0(\mathbf{x}) = -e^{\phi(\mathbf{x})}$ and the function $h(x) = -e^{x}$ is monotonically decreasing in $x$, $f_0$ is Schur-concave \cite[3.B.1]{Marshall1979Inequalities}.
	
	\bibliographystyle{IEEEtran}
	\bibliography{mybibliography.bib}

\begin{thebibliography}{10}
\providecommand{\url}[1]{#1}
\csname url@samestyle\endcsname
\providecommand{\newblock}{\relax}
\providecommand{\bibinfo}[2]{#2}
\providecommand{\BIBentrySTDinterwordspacing}{\spaceskip=0pt\relax}
\providecommand{\BIBentryALTinterwordstretchfactor}{4}
\providecommand{\BIBentryALTinterwordspacing}{\spaceskip=\fontdimen2\font plus
\BIBentryALTinterwordstretchfactor\fontdimen3\font minus
  \fontdimen4\font\relax}
\providecommand{\BIBforeignlanguage}[2]{{%
\expandafter\ifx\csname l@#1\endcsname\relax
\typeout{** WARNING: IEEEtran.bst: No hyphenation pattern has been}%
\typeout{** loaded for the language `#1'. Using the pattern for}%
\typeout{** the default language instead.}%
\else
\language=\csname l@#1\endcsname
\fi
#2}}
\providecommand{\BIBdecl}{\relax}
\BIBdecl

\bibitem{Yan2025Robust}
H.~Yan, B.~Song, A.~Ashikhmin, H.~Yang, and S.~Sun, ``{Robust Precoding for
  Massive MIMO Multi-Stream Multi-Satellite Systems},'' in \emph{2025 IEEE
  International Conference on Communications (ICC) Workshops, Montreal,
  Canada}, 2025.

\bibitem{ChengXiang2023Road}
C.-X. Wang, X.~You, and et~al., ``{On the Road to 6G: Visions, Requirements,
  Key Technologies, and Testbeds},'' \emph{IEEE Communications Surveys \&
  Tutorials}, vol.~25, no.~2, pp. 905--974, 2023.

\bibitem{38.811_3GPP}
3GPP, ``{Study on New Radio to Support Non-Terrestrial Networks (Release
  15)},'' \emph{document TR 38.811, V15.4}, Sep. 2020.

\bibitem{Jamshed2025Tutorial}
M.~A. Jamshed \emph{et~al.}, \emph{{A Tutorial on Non-Terrestrial Networks:
  Towards Global and Ubiquitous 6G Connectivity}}.\hskip 1em plus 0.5em minus
  0.4em\relax Now Foundations and Trends, 2025.

\bibitem{Qu2017LEO}
Z.~Qu, G.~Zhang, and et~al., ``{LEO Satellite Constellation for Internet of
  Things},'' \emph{IEEE Access}, vol.~5, pp. 18\,391--18\,401, 2017.

\bibitem{Di2019Ultra}
B.~Di, L.~Song, and et~al., ``{Ultra-Dense LEO: Integration of Satellite Access
  Networks into 5G and Beyond},'' \emph{IEEE Wireless Communications}, vol.~26,
  no.~2, pp. 62--69, 2019.

\bibitem{DELPORTILLO2019Comparison}
I.~{del Portillo}, B.~G. Cameron, and et~al., ``{A Technical Comparison of
  Three Low Earth Orbit Satellite Sonstellation Systems to Provide Global
  Broadband},'' \emph{Acta Astronautica}, vol. 159, pp. 123--135, 2019.

\bibitem{Marzetta2010NonCo}
T.~L. Marzetta, ``{Noncooperative Cellular Wireless with Unlimited Numbers of
  Base Station Antennas},'' \emph{IEEE Transactions on Wireless
  Communications}, vol.~9, no.~11, pp. 3590--3600, 2010.

\bibitem{Larsson2014mMIMO}
E.~G. Larsson, O.~Edfors, and et~al., ``{Massive MIMO for Next Generation
  Wireless Systems},'' \emph{IEEE Communications Magazine}, vol.~52, no.~2, pp.
  186--195, 2014.

\bibitem{Marzetta2016Fundamentals}
T.~L. Marzetta, E.~G. Larsson, H.~Yang, and H.~Q. Ngo, \emph{{Fundamentals of
  Massive MIMO}}.\hskip 1em plus 0.5em minus 0.4em\relax Cambridge, U.K.:
  Cambridge University Press, 2016.

\bibitem{You2020mMIMO}
L.~You, K.-X. Li, and et~al., ``{Massive MIMO Transmission for LEO Satellite
  Communications},'' \emph{IEEE Journal on Selected Areas in Communications},
  vol.~38, no.~8, pp. 1851--1865, 2020.

\bibitem{Ke-Xin2022Downlink}
K.-X. Li, L.~You, and et~al., ``{Downlink Transmit Design for Massive MIMO LEO
  Satellite Communications},'' \emph{IEEE Transactions on Communications},
  vol.~70, no.~2, pp. 1014--1028, 2022.

\bibitem{Xiang2024Massive}
Z.~Xiang, X.~Gao, K.-X. Li, and X.-G. Xia, ``{Massive MIMO Downlink
  Transmission for Multiple LEO Satellite Communication},'' \emph{IEEE
  Transactions on Communications}, vol.~72, no.~6, pp. 3352--3364, 2024.

\bibitem{Ziyu2026Decoupled}
Z.~Xiang, D.~Shi, R.~Sun, F.~Zhu, X.~Gao, and X.-G. Xia, ``{Decoupled Precoder
  and Receiver Design for Massive MIMO Multiple LEO Satellite Communication},''
  \emph{IEEE Transactions on Wireless Communications}, vol.~25, pp. 9747--9764,
  2026.

\bibitem{Zhao2023Robust}
B.~Zhao, M.~Lin, M.~Cheng, J.-B. Wang, J.~Cheng, and M.-S. Alouini, ``{Robust
  Downlink Transmission Design in IRS-Assisted Cognitive Satellite and
  Terrestrial Networks},'' \emph{IEEE Journal on Selected Areas in
  Communications}, vol.~41, no.~8, pp. 2514--2529, 2023.

\bibitem{Zhang2024Multi}
X.~Zhang, S.~Sun, and et~al., ``{Multi-Satellite Cooperative Networks: Joint
  Hybrid Beamforming and User Scheduling Design},'' \emph{IEEE Transactions on
  Wireless Communications}, vol.~23, no.~7, pp. 7938--7952, 2024.

\bibitem{Guidotti2024Federated}
A.~Guidotti, A.~Vanelli-Coralli, and C.~Amatetti, ``{Federated Cell-Free MIMO
  in Nonterrestrial Networks: Architectures and Performance},'' \emph{IEEE
  Transactions on Aerospace and Electronic Systems}, vol.~60, no.~3, pp.
  3319--3347, 2024.

\bibitem{Abdelsadek2023Broadband}
M.~Y. Abdelsadek, G.~Karabulut-Kurt, and et~al., ``{Broadband Connectivity for
  Handheld Devices via LEO Satellites: Is Distributed Massive MIMO the
  Answer?}'' \emph{IEEE Open Journal of the Communications Society}, vol.~4,
  pp. 713--726, 2023.

\bibitem{Yan2019Asyn}
H.~Yan and I.-T. Lu, ``{Asynchronous Reception Effects on Distributed Massive
  MIMO-OFDM System},'' \emph{IEEE Transactions on Communications}, vol.~67,
  no.~7, pp. 4782--4794, 2019.

\bibitem{Yafei2026Statistical}
Y.~Wang, V.~N. Ha, K.~Ntontin, H.~Yan, W.~Wang, S.~Chatzinotas, and
  B.~Ottersten, ``{Statistical CSI-Based Distributed Precoding Design for
  OFDM-Cooperative Multi-Satellite Systems},'' \emph{IEEE Journal on Selected
  Areas in Communications}, vol.~44, pp. 3219--3236, 2026.

\bibitem{Chaudhry2021Free}
A.~U. Chaudhry and H.~Yanikomeroglu, ``{Free Space Optics for Next-Generation
  Satellite Networks},'' \emph{IEEE Consumer Electronics Magazine}, vol.~10,
  no.~6, pp. 21--31, 2021.

\bibitem{Abdi2001estimation}
A.~Abdi, C.~Tepedelenlioglu, M.~Kaveh, and G.~Giannakis, ``{On the Estimation
  of the K Parameter for the Rice Fading Distribution},'' \emph{IEEE
  Communications Letters}, vol.~5, no.~3, pp. 92--94, 2001.

\bibitem{Ke-Xin2024Ergodic}
K.-X. Li, X.~Gao, and X.-G. Xia, ``{Ergodic Sum Rate Capacity Achieving
  Transmit Design for Massive MIMO LEO Satellite Uplink Transmission},''
  \emph{IEEE Transactions on Aerospace and Electronic Systems}, vol.~60, no.~3,
  pp. 3403--3416, 2024.

\bibitem{38.214_3GPP}
3GPP, ``{Physical Layer Procedures for Data (Release 19)},'' \emph{document TS
  38.214, V19.2}, Dec. 2025.

\bibitem{kailath2000linear}
T.~Kailath, A.~H. Sayed, and B.~Hassibi, \emph{{Linear Estimation}}.\hskip 1em
  plus 0.5em minus 0.4em\relax Upper Saddle River, NJ, USA: Prentice Hall,
  2000.

\bibitem{Shi2011WMMSE}
Q.~Shi, M.~Razaviyayn, Z.-Q. Luo, and C.~He, ``{An Iteratively Weighted MMSE
  Approach to Distributed Sum-Utility Maximization for a MIMO Interfering
  Broadcast Channel},'' \emph{IEEE Transactions on Signal Processing}, vol.~59,
  no.~9, pp. 4331--4340, 2011.

\bibitem{Boyd_Vandenberghe_2004}
S.~Boyd and L.~Vandenberghe, \emph{Convex Optimization}.\hskip 1em plus 0.5em
  minus 0.4em\relax Cambridge, U.K.: Cambridge University Press, 2004.

\bibitem{lanczos1950iteration}
C.~Lanczos, ``{An Iteration Method for the Solution of the Eigenvalue Problem
  of Linear Differential and Integral Operators},'' \emph{Journal of Research
  of the National Bureau of Standards}, vol.~45, no.~4, pp. 255--282, 1950.

\bibitem{Xin2024Asyncrhonus}
X.~Chen and Z.~Luo, ``{Asynchronous Interference Mitigation for LEO
  Multi-Satellite Cooperative Systems},'' \emph{IEEE Transactions on Wireless
  Communications}, vol.~23, no.~10, pp. 14\,956--14\,971, 2024.

\bibitem{Marshall1979Inequalities}
A.~W. Marshall and I.~Olkin, \emph{{Inequalities: Theory of Majorization and
  Its Applications}}.\hskip 1em plus 0.5em minus 0.4em\relax New York:
  Academic, 1979.

\end{thebibliography}

\end{document}